\documentclass[conference,10pt]{IEEEtran}
\IEEEoverridecommandlockouts

\usepackage{cite,graphicx,psfrag,amsmath,amsfonts}
\usepackage{amsthm,balance,verbatim,multicol}
\usepackage{subfigure,multirow,booktabs,placeins,float,todonotes}
\usepackage{xcolor,soul}
\usepackage{steinmetz,relsize}
\usepackage{enumitem,mathrsfs}
\usepackage{tabularx}
\usepackage{algorithm}
\usepackage[noend]{algpseudocode}

\DeclareMathOperator*{\argmin}{arg\,min}

\begin{document}

\title{Improving RF-DNA Fingerprinting Performance In An Indoor Multipath Environment Using Semi-Supervised Learning\vspace{-4.5mm}
\thanks{${\dagger}$ -- Corresponding author.} \thanks{This work supported in part by the University of Chattanooga (UC) Foundation, Inc.}
}

\author{
  \IEEEauthorblockA{Mohamed K. M. Fadul, Donald R. Reising$^{\dagger}$, Lakmali P. Weerasena, T. Daniel Loveless, Mina Sartipi, Joshua H. Tyler$^{\ast}$} \vspace{-4.5mm} \\
  \normalsize The University of Tennessee at Chattanooga  \\
  \normalsize Chattanooga, TN 37403 USA \\
  \normalsize Email:\{mohammed-fadul, donald-reising$^{\dagger}$, lakmali-weerasena, daniel-loveless, mina-sartipi\}@utc.edu, ygm111@mocs.utc.edu$^{\ast}$ \\
}

% make the title area
\maketitle

\begin{abstract}
The number of Internet of Things (IoT) deployments is expected to reach 75.4 billion by 2025. Roughly 70\% of all IoT devices employ weak or no encryption; thus, putting them and their connected infrastructure at risk of attack by devices that are wrongly authenticated or not authenticated at all. A physical layer-based security approach--known as Specific Emitter Identification (SEI)--has been proposed and is being pursued as a viable IoT security mechanism. SEI is advantageous because it is a passive technique that exploits inherent and distinct features that are unintentionally imparted upon the signal during its formation and transmission within and by the IoT device's Radio Frequency (RF) front-end. SEI's passive exploitation of unintentional signal features removes any need to modify the IoT device, which makes it ideal for existing and future IoT deployments. Despite the amount of SEI research conducted there remains challenges that must be addressed to make SEI a viable IoT security approach. One of these challenges is the extraction of SEI features from signals collected under multipath fading conditions. Multipath corrupts the inherent SEI exploited features that are used to discriminate one IoT device from another; thus, degrading authentication performance and increasing the chance of attack. This work presents two semi-supervised Deep Learning (DL) equalization approaches and compares their performance with the current state of the art. The two approaches are the Conditional Generative Adversarial Network (CGAN) and Joint Convolutional Auto-Encoder and Convolutional Neural Network (JCAECNN). Both approaches learn the channel distribution to enable multipath correction while simultaneously preserving the SEI exploited signal features. CGAN and JCAECNN performance is assessed using a Rayleigh fading channel under degrading SNR, up to thirty-two IoT devices, as well as two publicly available signal sets. The JCAECNN improves SEI performance by 10\% beyond that of the current state of the art.
\end{abstract}
\section{Introduction}%
\label{sect:introduction}
\IEEEPARstart{I}{\lowercase{t}} is estimated that the number of operational Internet of Things (IoT) devices will reach 75.4 billion by 2025~\cite{Maayan_IoT_2020}. Disturbingly, roughly seventy-percent of IoT devices employ weak or no encryption at all due to limited onboard resources (e.g., memory, computation, etc.), manufacturing costs that prohibit adoption, or difficulties associated with key management and implementation at scale~\cite{Rawlinson_2014,Ray_CIC_2019,Neshenko_CommSurvey_2019}. The lack of or limited encryption creates a security vulnerability that leaves individual IoT devices and corresponding infrastructure open to exploitation by nefarious actors~\cite{Larsen_CNN_2017,Wright_book_2015,Stanislav_Rapid_2015,Wright_Killer_2019,Simon_2016,Shipley_DEFCON_2014,Shipley_GitHub,Krebs_Mirai_2017}. Based upon this observation and attacks there is a critical need for an effective IoT security solution. 
%One such solution is the physical layer approach of Specific Emitter Identification (SEI)~\cite{Talbot_CandS_2017,Sa_Access_2019,Reising_IoT_2020}.
\\
\indent IoT deployments conform to the Open Systems Interconnection (OSI) model, which states that the %According to the Open Systems Interconnection (OSI) model, the 
Physical (PHY) layer is responsible for completing point-to-point bit stream communication and is considered the lowest and first layer~\cite{OSI}. By design, traditional security techniques are implemented at higher OSI layers (e.g., Data Link or Network)% such techniques include: encryption, IP and MAC address filtering
; thus, they %these security techniques 
do not consider the PHY layer despite the fact that attackers must traverse it to conduct their attacks% it is the first layer traversed by attackers%exposed to attacks
~\cite{Reising_Dissertation}. Specific Emitter Identification (SEI) has been proposed as a PHY layer IoT security solution%is a PHY layer technique being put % has been put forward %as a viable means through which to secure IoT devices and their associated infrastructure% at the PHY layer
~\cite{Talbot_CandS_2017,Sa_Access_2019,Reising_IoT_2020}. SEI has been shown capable of achieving serial number discrimination by exploiting immutable, unintentional coloration that is imparted upon a signal during its formation and transmission. The coloration's source is attributed to the tolerated, manufacturing variation(s) present within the individual components, sub-systems, and systems that comprise an emitter's Radio Frequency (RF) front-end%SEI exploits %identifies a wireless device by exploiting %s the immutable, unintentional coloration--that is %unintentionallyimparted upon every signal during its formation and transmission. The components, sub-systems, and systems that comprise a transmitter's Radio Frequency (RF) front-end% to identify wireless devices
~\cite{Jeffery_MobiCom_2007,DudczykSEI2,BrikMobi08,Suski_IJESDF_2008,DanevIPSN09,Klein_ICC_2009,Liu_SEI_2011,Kennedy_2010,Reising_Dissertation,Williams_NSS_2010,Takahashi_CompApps_2010,TekbasIEE,Ellis_RadioSci,Soli_IEEE,CanadaSEI,azzouz,Reising_InfoSec_2015,Wheeler_ICNC_2017,Pan_2019, Fadul_InfoSec_2020, Fadul_WCNC_2019,Fadul_GIoTS_2022}. SEI is advantageous because it provides a passive authentication mechanism that does not require modification of the end device (i.e., the coloration is generated as part of normal operations), which makes it ideally suited to IoT applications since there is no additional resource demands placed on the device being identified. However, since its inception--almost thirty years ago--very little attention has been paid to the performance of SEI within contentious operating conditions such as multipath. SEI's viability--as an IoT security solution--rests on it remaining effective even under degrading operating conditions.
%more advanced authentication mechanism than traditional OSI higher layer (e.g., data link and network layer) techniques that become vulnerable to exploitation especially when no encryption is used. ... Need to address/integrate.

%
\begin{table*}[!t]
\vspace*{-0.05in}
\renewcommand{\arraystretch}{1.3}
  \caption{Notations.% \hl{Mohamed, update this table to address all of notations you used in this article.}
  %\todo{I would look at the order in which things are presented in Table 1. $I_{A}$ is defined by $A$, but introduced before A. Same thing with $X_{e}$ and R. Also, is $r[n]$ and $x[n]$ supposed to be $x(t)$? $z$ has not been defined, but is used to define $G(z)$}
  }
  \label{tbl:notation}
  \centering
  \begin{tabular}{|c|m{.23\textwidth}||c|m{.23\textwidth}||c|m{.23\textwidth}|}
  \hline
    $N_{D}$ & Number of emitters used in this work &
    $N_{B}$ & Number of collected signals per emitter &
    $A$ , $B$ & I.I.D random variables  \\
    $\sigma^{2}$ & Variance & %
    $L$ & Length of Rayleigh fading channel &
    $T_{r}$ & Root-Mean-Squared of the delay spread \\
    $T_{s}$ & Sampling period &
    $\alpha_{k}$ & Coefficient of the $k^{th}$ Rayleigh fading path &
    $\tau_{k}$ &  Delay of the $k^{th}$ Rayleigh fading path \\
    $x(t)$ & Collected IEEE 802.11a signal &
    $r(t)$ & Received signal with multipath &
    $n(t)$ & Complex, white Gaussian noise \\
    $h(t)$ & TDL channel impulse response &
    $\hat{x}(m)$ & Estimate of the transmitted signal $x(m)$ &          %%% start here
    $\gamma$ & Signal-to-Noise ratio (SNR) \\
    $I_{A}$ & Identity matrix with matching dimensions to $A$ &
    $\beta$ & NN activation function &
    $W$ & Weights of the convolutional kernel \\
    $\mathbf{\mathcal{F}}$ & Set of RF-DNA fingerprints with class labels &
    $b$ & NN bias &
    $\theta$ & Trainable parameters of the NN \\
    $J(\theta)$ & Loss function of the NN &
    $D(x)$ & Discriminator's output for input $x$ & 
    $G(z)$ & Generator's output for input $z$ \\
    $P_{d}(x)$ & Distribution of training data $x$ & 
    $P_{z}(z)$ & Prior probability of $G$'s input $z$ &
    $r_{\mathscr{R}}(n)$ & Received signal In-phase component \\
    $r_{\mathscr{I}}(n)$ & Received signal Quadrature component &
    $\rho_{r}$ & Magnitude of the received signal &
    $\theta_{r}$ & Phase angle of the received signal \\
    $N_{R}$ & Number of training signals per emitter &
    $N_{T}$ & Number of testing signals per emitter &
    $\tilde{x}[n]$ & AWGN channel output for input $x[n]$ \\
    $y$ & Actual Label corresponding to the transmitted signal $x[n]$ &
    $\tilde{X}_{e}$ & IQ+NL labeled representation of $\tilde{x}[n]$ &
    $\mathbf{R}_{e}$ & IQ+NL labeled representation of $r[n]$  \\
    $\hat{X}_{e}$ & $G$'s output for input $\mathbf{R}_{e}$ &
    $\tilde{y}_{c}$ & Estimated label by the JCAECNN classifier &
    $C_{k}$ & Delayed and scaled version of the trasnmitted signal $x[n]$  \\
    $N_{z}$ & Number of noise realizations & $z$ & Input to the CGAN generator, $G$  &  & \\
    \hline
  \end{tabular}
  \vspace{-4mm}
\end{table*}

RF-Distinct Native Attributes (RF-DNA) fingerprinting is an SEI implementation that extracts exploitable features from portions of the transmitted signal corresponding to fixed, known symbol sequences such as the IEEE 802.11a Wireless-Fidelity (Wi-Fi) preamble. Prior RF-DNA fingerprinting research %presents traditional implementation of the RF-DNA fingerprinting which 
uses supervised and unsupervised learning algorithms that rely on expert knowledge and handcrafted features to facilitate IoT device discrimination%discriminate between IoT devices (emitters)
~\cite{Suski_IJESDF_2008,Klein_ICC_2009,Williams_NSS_2010,Wheeler_ICNC_2017,Liu_SEI_2011,Kennedy_2010,Reising_Dissertation,Fadul_WCNC_2019,Fadul_thesis,Pan_2019,KandahiThings2019,Fadul_InfoSec_2020,Reising_IoT_2020,Fadul_Access_2021,Fadul_GIoTS_2022}. These algorithms may results in sub-optimal models that limit RF-DNA fingerprinting performance, especially under degraded operating conditions such as %are changing due to 
time-varying or lower Signal-to-Noise Ratio (SNR) channels. Over the past five years the SEI research community has pursued %investigated the use of 
Deep Learning (DL) due to its successful application in %Due to the demonstrated success of Deep Learning (DL) in 
spectrum management \cite{DARPA_SC2,Yu_ICC_2018}, modulation and emitter identification \cite{Shea_CR_2017,DARPA_RFMLS,Restuccia_DeepRadioID_2019}, system design \cite{Shea_CR_2017,Restuccia_Mobihoc_2020,Qin_WComms_2019,Downey_Spectrum_2020,Fadul_Milcom_21}, as well as its ability to thrive under increasing amounts of information while removing the need for handcrafted features%. Recent efforts have proposed DL-based implementations of SEI and RF-DNA fingerprinting
~\cite{Baldini_interference_2019, Baldini_comparison_2019, Shea_Air_2019, wong_clustering_2018, Gihan_hardware_2019, Riyaz_CNN_2018, Merchant_DL_fingerprinting_2018, Restuccia_DeepRadioID_2019, Jafari,Guyue_Wifi_2019, Youssef_ML_2018,Pan_2019,Yu_WiMob_2019,Jian_Massive_study_2020,Robinson_Dilated_2020, Ding_IET_2018, Chen_IET_2020, Peng_TVT_2020, Yang_ICT_2021, Morin_SIP_2019, Behura_TCCN_2020, McGinthy_IoT_2019, Tang_2021, Liu_IoT_2021, Zha_ICT_2021, Gong_WCSP_2019, Ji_AUTEEE_2020, Gong_infosec_2020, Li_AIID_2021, Wang_Comm_2021, Qu_Sym_2021, Bassey_FMEC_2019, Wang_ICICSP_2021, Cun_ICCT_2021, Peng_CCC_2021, Shen_CIS_2018,Tyler_ICC_2022}. Based upon the success of these published DL works, our work--presented herein--shows that DL can provide an effective RF-DNA fingerprinting process capable of facilitating serial number discrimination of IoT devices under degraded operating conditions. Specifically, we show that up to thirty-two IoT devices--that only differ in serial number--can be successfully discriminated from one another within an indoor multipath channel environment.

The remainder of this paper is organized as follows. Sect.~\ref{sect:movtivation} summarizes current publications that are most pertinent to our work as well as providing a more detailed description of our work's contributions. Sect.~\ref{sect:Background} provides descriptions of the; signal of interest, signal collection and detection processes, multipath channel modeling, Nelder-Mead (N-M) channel estimator and Minimum Mean Square Error (MMSE) equalizer, as well as the deep learning architectures used herein. The methodology for the developed DL-driven RF-DNA fingerprinting approach for indoor multipath environments is described in Sect.~\ref{sect:methodology}. The results are presented in Sect.~\ref{sect:Results} and the article concluded in Sect.~\ref{sect:SumConc}.
\section{Related Work and Contribution}%
\label{sect:movtivation}
DL-based SEI has been shown to provide an optimal end-to-end solution that results in superior performance, however most of the published works do not conduct SEI using RF signals that have undergone time-varying fading. %~\cite{Gihan_hardware_2019,wong_clustering_2018,Jafari,Riyaz_CNN_2018,Guyue_Wifi_2019,Liu_IoT_2021,Qu_Sym_2021,Shen_CIS_2018, Baldini_interference_2019,Merchant_DL_fingerprinting_2018,Youssef_ML_2018,Baldini_comparison_2019,Yu_WiMob_2019,Ding_IET_2018,Peng_TVT_2020,Yang_ICT_2021,McGinthy_IoT_2019,Tang_2021,Zha_ICT_2021,Ji_AUTEEE_2020,Li_AIID_2021,Wang_Comm_2021,Bassey_FMEC_2019,Cun_ICCT_2021,Peng_CCC_2021}. 
In fact, only a small subset of DL-based SEI works have assessed its performance under multipath conditions and most can be categorized as using a channel that is: (i) static (i.e., the channel coefficients or characteristics never change)~\cite{Pan_2019}, or (ii) time-varying, but the environments specific characteristics (e.g., number of reflectors, line-of-sight present or not, type of fading, etc.) are unknown, unstated, or not disclosed by the authors~\cite{Pan_2019,Shea_Air_2019,Restuccia_DeepRadioID_2019,Jian_Massive_study_2020,Robinson_Dilated_2020,Gong_infosec_2020,Gong_WCSP_2019,Wang_ICICSP_2021}. %The remainder of this section summarizes the contributions of these works and states the contributions of our work presented herein.
Regardless, most DL-based SEI efforts assume %The majority of these efforts assume 
the selected, modified, or developed %DL
algorithm can learn discriminatory features directly from the received signals; thus, eliminating the need for channel estimation and correction prior to RF signal classification. The fact is that the time-varying nature of multipath fading channels impedes or significantly hinders the DL algorithms' ability to learn discriminating signal features that are invariant to multipath fading.

The work in~\cite{Fadul_Access_2021} is the exception to this trend in that the authors: (i) adopt the IEEE 802.11a Wi-Fi indoor, channel model and state the specific values used to configure it~\cite{Ohara_Book_2005} as well as (ii) perform traditional (i.e., not DL-based) multipath channel estimation and correction prior to performing SEI using a Convolutional Auto-Encoder (CAE) initialized Convolutional Neural Network (CNN). The authors' use of traditional estimation and correction approaches results in sub-optimal SEI performance due to errors in the estimated channel coefficients and a dependence on knowing the channels' statistics. Lastly, the authors assess their CAE-CNN SEI approach using only four emitters, which does not reflect typical IoT deployments consisting of tens to hundreds of devices.%\todo{Moe, Did I miss anything?}

Our work--presented in detail in the remainder of this paper--advances DL-based SEI's current state of the art through the following contributions:

\begin{enumerate}[leftmargin=*]
\item{Combining label embedding with a Conditional Generative Adversarial Network (CGAN) to %representation to of %in~\cite{GlobeCom_Tyler_2021} to represent 
efficiently learn each emitter's conditional feature distribution.%s of each emitter the IQ samples to enable generative models given by CGANs to  efficiently learn conditional feature distributions for all emitters.}
}
\item{Augmenting discriminatory feature learning through the use of the rectangular and polar signal representation~\cite{GlobeCom_Tyler_2021}.}
\item{Introduces and assesses two semi-supervised learning equalization approaches that facilitate superior CNN-based SEI under Rayleigh fading and degrading SNR conditions. %prior to CNN classification to replace the traditional channel estimation in~\cite{Fadul_Access_2021}, where the channel equalization/correction is performed using semi-supervised learning-based algorithms. The DL based RF-DNA fingerprinting performance is assessed using a Rayleigh fading channels under degrading SNR.
}
\item{Introduces and assesses a scalable, semi-supervised learning architecture that jointly develops CAE-based generative models %--given by CAEs--and 2) 
and a CNN-based discriminative model % given by a CNN classifier in a joint architecture to 
to correct for Rayleigh fading while preserving SEI discriminative features. This joint architecture--designated herein as JCAECNN--%approach trains a generative model to 
decomposes the multipath signal into individual scaled and delayed versions of the original transmitted signal prior to CNN classification.%by a CNN network.
}
\item{Improving our JCAECNN's SEI performance through the use of exponentially decaying loss function weights.}%of the proposed joint CAE and CNN (JCAECNN) architecture by using exponentially decaying loss function weights.}}
%\item{Optimization using exponentially decaying loss function weights}

%\item{The Use of Conditional Generative Adversarial Network (CGAN) to provide Multipath channel correction. The CGAN learns the Rayleigh distribution of the multipath IEEE 802.11a Wi-Fi signals and estimate a generative model that performs channel equalization.}

\item{SEI performance assessment of the CGAN and JCAECNN architectures as four, eight, sixteen, or thirty-two emitters communicate over a Rayleigh fading channel.%Scalability analysis of the CGAN and JCAECNN %two proposed semi-supervised SEI approaches by assessing their performance under Rayleigh fading channel when 4, 8, 16, and 32 emitters are used.
% Introduction of a scalable semi-supervised learning-based approach that uses a Joint CAE and CNN to correct for Rayleigh channel effects. This approach trains a generative model to decompose the multipath signal to individual scaled and delayed versions of the original transmitted signal prior to classification by a CNN network.
}

\item{JCAECNN SEI performance assessment using the public signal sets generated by the authors of~\cite{Oracle_2019,WiSig}. These results serve as a benchmark to permit comparative assessment with current and future publications.%, publicly availaof the  architecture usingperformance for different datasets including two publicly available datasets collected by previous works in~\cite{Oracle_2019,WiSig}. This analysis provides a benchmark to facilitate comparative assessment against future works.
}
%\item{Channel invariance. % Optimize the loss function coefficient of Joint CAE and CNN system by assigning exponentially decaying values during training to improve the model training accuracy and overall SEI performance
%}

\item{Directly compares the CGAN and JCAECNN architectures' SEI performance with our prior work in~\cite{Fadul_Access_2021,Fadul_GIoTS_2022}.%In comparison to our previous work in~\cite{Fadul_Access_2021,Fadul_GIoTS_2022}, the use of semi-supervised channel equalization improves the SEI identification performance by  Improving the SEI performance accuracy by an average of 10\% at SNR$=$9~{dB}.
}
\end{enumerate}

These contributions result in an average percent correct classification performance of 94.35\% or better for SNR values of $9$~{dB} or higher, Rayleigh fading channels comprised of five reflections/paths and an IoT deployment consisting of sixteen devices.
\section{Background%
\label{sect:Background}}
\subsection{Signal of Interest%
\label{sect:Signal_of_Interest}}
\begin{figure}[!b]
  \centering
  \vspace{-5mm}
  \includegraphics[width=\columnwidth]{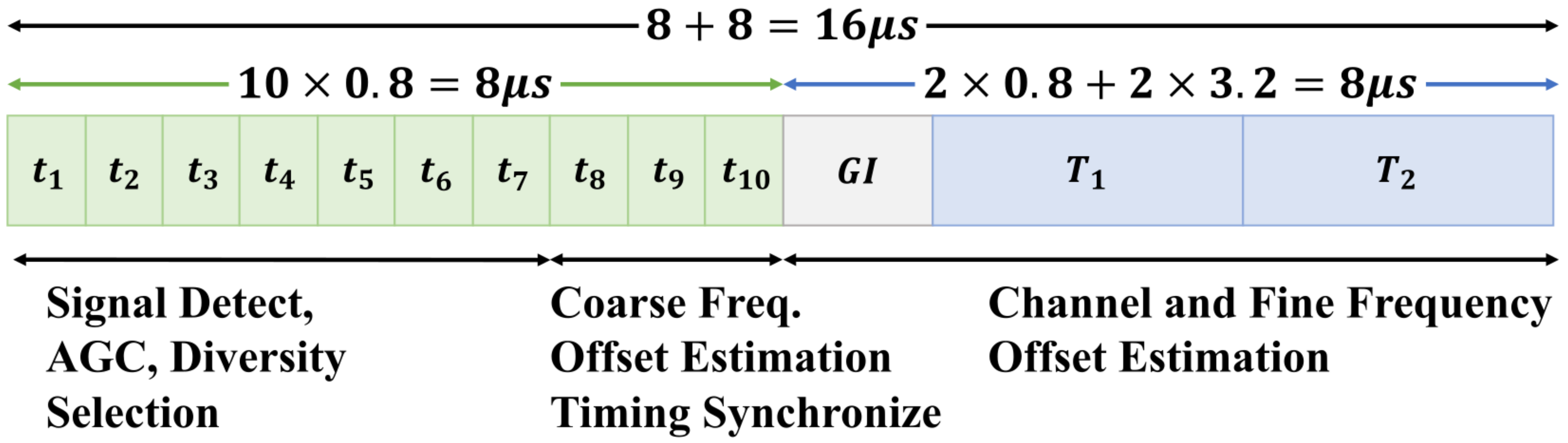}
    \caption{Structure of the 16~{$\mu$s} duration preamble that is present at the start of every IEEE 802.11a Wi-Fi frame~\cite{802.11}.}
    % \vspace{-3mm}
    \label{fig:preamble_structure}
\end{figure}

This work makes %The work presented here makes 
use of IoT emitters that communicate using the IEEE 802.11a Wi-Fi protocol. The reasons for choosing the %Our choice of using 
IEEE 802.11a protocol for the signal sub-layer are as follows: %is for the based on three factors: 1) it is an IoT designated communications protocol~\cite{wisilica_2020}
(i) 802.11a is based on Orthogonal Frequency Division Multiplexing (OFDM) signal, which is used in multiple wireless communication standards such as 802.11ac, 802.11ad, 802.11ax, Long Term Evolution (LTE), and Worldwide Interoperability for Microwave Access (WiMAX)~\cite{Lajos_2011}, (ii) multiple research efforts within the SEI community have demonstrated success using 802.11a signals~\cite{Jeffery_MobiCom_2007,Suski_IJESDF_2008,Liu_SEI_2009,Takahashi_CompApps_2010,Liu_SEI_2011,Reising_InfoSec_2015,Wheeler_ICNC_2017,Fadul_WCNC_2019,Fadul_thesis,Fadul_InfoSec_2020,Riyaz_CNN_2018,Restuccia_DeepRadioID_2019,Tyler_ICC_2022,GlobeCom_Tyler_2021,Fadul_GIoTS_2022}, (iii) availability of the same data set used in our previous publications~\cite{Fadul_Access_2021,Fadul_WCNC_2019,Fadul_InfoSec_2020,Tyler_ICC_2022,GlobeCom_Tyler_2021,Fadul_GIoTS_2022} to facilitate comparative assessments, and (iV) 802.11a Wi-Fi has been adopted as an IoT communications protocol~\cite{wisilica_2020}. Consistent with our prior publications, this work performs SEI by extracting RF-DNA fingerprints from the IEEE 802.11a Wi-Fi preamble, which comprises the first 16~$\mu$s of every transmission~\cite{Fadul_WCNC_2019,Fadul_Access_2021,Fadul_InfoSec_2020,Fadul_GIoTS_2022}. Use of the 802.11a preamble is ideal, because it is the portion of the signal used by the receiver to perform channel equalization. An 802.11a preamble consists of ten--designated $t_{1}$ through $t_{10}$--Short Training Symbols (STS), a Guard Interval (GI), and two Long Training Symbols (LTS) that are designated $T_{1}$ and $T_{2}$ %~\cite{802.11}. The structure of the IEEE 802.11a Wi-Fi preamble is shown 
in Fig.~\ref{fig:preamble_structure}~\cite{802.11}.
\subsection{Signal Collection \& Detection%
\label{sec:data_collection}}
The primary data set used in this work is comprised of 802.11a Wi-Fi signals transmitted by $N_{D}=4$ Cisco AIR-CB21G-A-K9 Wi-Fi emitters and collected using an Agilent E3238S-based spectrum analyzer. %from $N_{D}=4$ Cisco AIR-CB21G-A-K9 Wi-Fi emitters/emitters. 
The spectrum analyzer can sample signals at rates up to % sampling rate of up to
95~{mega-samples per second~(Msps)}, has an operating range from 20~{MHz} to 6~{GHz}, an RF bandwidth of 36~{MHz}, and a 12-{bit} analog-to-digital converter~\cite{RFSICS}. 
Upon collection, a total of $N_{B}=2,000$ signals are selected for each of the $N_{D}=4$  emitters using amplitude-based variance trajectory detection~\cite{Klein_ICC_2009}. After that, each signal is filtered %all signals from all emitters are filtered 
using a fourth order low-pass Butterworth filter with a cutoff frequency of 7.7~{MHz}. Following filtering, the signals are post-processed by correcting for carrier frequency offset and downsampled to %a sampling rate of 
20~{MHz}~\cite{Fadul_thesis}. 

\subsection{Multipath Channel Modeling%
\label{sect:Multipath_Model}}
Multipath is a major concern for communications systems operating in indoor environments. %one of the major concerns in indoor communications. 
It degrades %negatively affects the 
system performance by limiting the receiver's ability to correctly recover %decode (reconstruct) 
the transmitted message. This degradation is due to multiple copies of the transmitted signal--associated with different delays and attenuation values--destructively interfere with one another at the receiver's antenna(s). %at 802.11a Wi-Fi transceivers. 
These copies are reflections of the % are generated due to the reflection of the 
transmitted signal off of objects located throughout the propagation environment~\cite{Ohara_Book_2005}. The combination of these reflected copies--%echoes 
at the receiver--randomly shifts the carrier frequency between the transmitter and receiver as well as changes the signal strength %of the signal 
over short time intervals \cite{Ohara_Book_2005}. %~\todo{Is this citation correct? Double check it.}\cite{Ohara_Book_2005}.
For the work presented in this article, the indoor, multipath channels are modeled using a Rayleigh distribution for two reasons: (i) this distribution is used to assess Wi-Fi modulation performance by the IEEE 802.11 working group~\cite{Ohara_Book_2005}, and (ii) prior SEI publications have used or assessed system performance using the same distribution~\cite{Fadul_WCNC_2019,Fadul_InfoSec_2020,Fadul_Access_2021,Fadul_GIoTS_2022,Shea_Air_2019,Wang_ICICSP_2021,Gong_infosec_2020,Gong_WCSP_2019}.%\todo{Moe, Add the others}.%To assess the modulation performance within indoor multipath environments, IEEE 802.11 working group uses Rayleigh distribution to model the multipath channel \cite{Ohara_Book_2005}. Therefore, this work selects Rayleigh fading channel to capture the time-varying nature of the wireless multipath environment.  

In multipath fading, a single reflection (a.k.a., path) is quantified using %Each multipath reflection path is quantified by a
coefficient and a corresponding time delay known as a delay spread~\cite{Ohara_Book_2005}. For the case of Rayleigh fading, a Tap Delay Line (TDL) %A Tap Delay Line (TDL) 
is used to mathematically describe the channel. In a TDL, each ``tap'' %the Rayleigh fading channel, where each tap
represents a single coefficient and delay. For each path $k$, the coefficient is represented using a circularly symmetric complex Gaussian random variable,% as follows,
\begin{equation}
	\alpha_{k} = A + jB,
	\label{eqn:complex_coeff}
\end{equation}
where $k=1, \dots, L$ for a total number of $L$ paths comprising the channel, and $\sigma^{2}$ is the variance of the zero mean independent and identically distributed Gaussian random variables $A$ and $B$~\cite{Ohara_Book_2005}. If the  delay spread's Root-Mean-Squared (RMS) %of the delay spread 
is $T_{\text{r}}$ and the sampling period is $T_{s}$, then the variance can be defined as,
\begin{equation}
	\sigma^{2} = \dfrac{\sigma^{2}_{k}}{2} = \dfrac{1}{2}\left\{\left[1 - \exp\left(\dfrac{-T_{s}}{T_{\text{r}}}\right)\right]\exp\left(\dfrac{-kT_{s}}{T_{\text{r}}}\right)\right\}.
	\label{eqn:variance}
\end{equation}
Finally, the TDL for a %representing the 
multipath environment consisting of $L$ total reflecting paths is,
\begin{equation}
	h(t) = \sum\limits_{k=1}^{L}{\alpha_{k}\delta(t - \tau_{k}T_{s})},
	\label{eqn:tdl_eqn}
\end{equation}
where $\alpha_{k}$ and $\tau_{k}$ are the coefficient and delay spread associated with the $k^{\text{th}}$ path~\cite{Ohara_Book_2005,Hijazi_TransVehicle_2009}. A received signal with multipath is generated by,%For each IEEE 802.11a collected signal $x(t)$, the Rayleigh fading TDL equation~\eqref{eqn:tdl_eqn} is applied to generate a received multipath signal $r(t)$ as follows,
\begin{equation}
	r(t) = x(t) \ast h(t) + n(t),
	\label{eqn:rcvd_signal}
\end{equation}
where $x(t)$ is a collected 802.11a Wi-Fi signal, $h(t)$ is the TDL from~\eqref{eqn:tdl_eqn}, $n(t)$ is complex, white Gaussian noise, and $\ast$ denotes the convolution operation. The noise $n(t)$ is filtered, scaled, and added to the result of $x(t) \ast h(t)$ to generate multipath signals with SNR values from 9~{dB} to 30~{dB} in increments of 3~{dB} between consecutive values.   
%The variance of the \textbf{??filtered??}\todo{Moe, when do you scale the noise?} noise $n(t)$ is scaled to generate multipath signals with SNR values from 9~{dB} to 30~{dB} in increments of 3~{dB} between consecutive values.
%
\subsection{Traditional Channel Estimations \& Equalization}
\label{sec:trad_est}
This section explains the Nelder-Mead (N-M) estimator and MMSE equalizer, which are used here to facilitate comparative assessment between our results in Sect.~\ref{sect:Results} and those presented in our published work~\cite{Fadul_Access_2021}.

\begin{comment}
This section provides a background in channel estimation and equalization as a preprocessing step prior to applying DL model for feature extraction and classification. This preprocessing step has been applied by our previous work in~\cite{Fadul_Access_2021} to correct for multipath channel effects and improve the DL-based SEI performance.  
The receiver performs channel impulse response estimation to calculate the coefficient associated with each path \cite{Fadul_WCNC_2019}. Our previous work in\cite{Fadul_Access_2021} uses Nelder-Mead (N-M)~\cite{Fadul_WCNC_2019} channel estimator based on the comparative assessment under degrading SNR results of Nelder-Mead (N-M)~\cite{Fadul_WCNC_2019}, Least Square (LS)~\cite{Yuan_Inform_2008}, Minimum Mean Squared Error (MMSE)~\cite{Beek_1995,Edfors_1998}, and Adaptive Compensator (A-C)~\cite{Kennedy_2010} 
which shows that N-M estimator achieves superior performance to all the other algorithms \cite{Fadul_InfoSec_2020}.
\end{comment}

\subsubsection{Nelder-Mead Channel Estimator%
\label{sec:nm_estimator}}
The N-M estimator is constructed using the %N-M estimator is based on 
N-M simplex algorithm, which % presented in~\cite{Nelder_CompJrnl_1965}. N-M 
minimizes unconstrained optimization problems using a direct search approach~\cite{Nelder_CompJrnl_1965}. %attempts to find the minimum of unconstrained optimization problems. 
The N-M simplex algorithm iteratively determines a $d$-variable, non-linear function's minimum solution using %uses 
only function values and %along with 
four defined operations. %iteratively to find the minimum solution of a $d$-variable non-linear function. 
The N-M simplex algorithm is computationally efficient, because it does not require computation of the % of the N-M algorithms comes from the fact that it doesn't require computing of the  
function derivatives \cite{lagarias_Optim_2006}. At the start of iteration $j$, a $d+1$ vertices simplex is defined, one or more of four operations--reflection, expansion, contraction, and shrinkage--are performed to calculate one or more new points, and the function's values calculated using the new point(s). If the calculated function values at vertices given by $x_{1}$ through $x{d+1}$ satisfy certain conditions, then the new point(s) replace the worst point(s) in the simplex that is denoted as $x_{d+1}$~\cite{lagarias_Optim_2006, Nelder_CompJrnl_1965}. %is used to solve the optimization problem iteratively, with a simplex of $d+1$ vertices defined at the beginning of each iteration $j$. Within each iteration, one or more of the four operations named: reflection, expansion, contraction, and shrinkage is performed to calculate one or more points (vertices). When the function values at the new points calculated by reflection, expansion, and contraction satisfy certain conditions detailed in~\cite{lagarias_Optim_2006, Nelder_CompJrnl_1965}, the new point replaces the worst point in the simplex denoted by $x_{d+1}$. 
If none of the first three operations' conditions are %associated with the first three operations is 
satisfied, then a new set of points--$v{2}$ through $v{d+1}$--are calculated using the shrinkage operation %is used to calculate a new set of points $x(2)$ through $x(d+1)$
using the following formula ~\cite{lagarias_Optim_2006},
\begin{equation}
	v_{i}=x_{i}+\varphi(x_{i}-x_{1}),
	\label{eqn:Shrinkage}
\end{equation} 
where $1<i\leq d+1$, and the new simplex for the next iteration is $(x_{1}, v_{2}, \cdots v_{d+1})$~\cite{lagarias_Optim_2006}. The N-M simplex algorithm stops when the calculated function value--at iteration $j$--satisfies a certain termination condition or conditions. The termination conditions used in estimating the Rayleigh fading channel's coefficients are the same as those used in~\cite{Fadul_WCNC_2019, Fadul_InfoSec_2020,Fadul_GIoTS_2022}, which are given by the following formulas, %...\todo{Moe, add the conditions here.} 
\begin{equation}
	\dfrac{1}{d}\sum\limits_{i=1}^{d+1}{[f(x_{i}) - \bar{f}]^{2} < \epsilon_{1}},
	\label{eqn:condition_one}
\end{equation}
\begin{equation}
	\dfrac{1}{d}\sum\limits_{i=1}^{d}{\left\|x_{i}^{j} - x_{i}^{j+1}\right\|^{2} < \epsilon_{2}},
	\label{eqn:condition_two}
\end{equation}
where $\bar{f}$  is the average of the function values $f(x_{i})$, $\|\bullet\|$ is the $l_{2}$-norm, and $\epsilon_{1}$ and $\epsilon_{2}$ are tolerances based on the function values $f(x_{i})$ and points $x_{i}$ respectively. Both conditions are checked at the end of each iteration.
%The two stopping criteria based on the function values and the vertices themselves used in~\cite{Fadul_WCNC_2019, Fadul_InfoSec_2020,Fadul_GIoTS_2022} are adopted as termination criteria in this work.
The function to  be minimized is, %The function--to be minimized by the N-MTo apply N-M algorithm to estimate the Rayleigh fading channel coefficients, the function to be minimized can be formulated by,
\begin{equation}
	f(h) = \mathlarger{\sum}_{m \in T}{\bigg|r(m) - \sum\limits_{k=1}^{L}{x(m - \tau_{k})\alpha_{k}\bigg|^{2}}}.
	\label{eqn:to_be_minimized}
\end{equation}
The function $f(h)$ can be described as a square error function between the received signal and the weighted and delayed copies of the transmitted signal $x(m)$~\cite{Fadul_WCNC_2019}. Equation \eqref{eqn:to_be_minimized} is minimized in two parts because $r(m)$ and $x(m)$ are complex-valued while the N-M simplex algorithm solves real-valued functions \cite{lagarias_Optim_2006,Nelder_CompJrnl_1965}. %The solution to equation \eqref{eqn:to_be_minimized} is done in two parts due to the fact that both $r(m)$ and $x(m)$ are complex-valued while N-M algorithm is designed to solve real-valued functions \cite{lagarias_Optim_2006,Nelder_CompJrnl_1965}. 
The two parts are created by separating the complex-valued signals into their real and imaginary components; thus, resulting in % formulated by grouping the real and imaginary parts to create 
two real-valued, square error functions~\cite{Fadul_InfoSec_2020}. The N-M simplex algorithm is applied to each function separately to calculate the coefficient values $\alpha_{k}$ and the the estimation error reduced by using the average of the estimates from the two square error functions as the final coefficient estimates%the final coefficient estimates are calculated as the average of the corresponding estimates from the two square error functions 
~\cite{Fadul_InfoSec_2020}.
As in~\cite{Fadul_WCNC_2019, Fadul_InfoSec_2020,Fadul_Access_2021,Fadul_GIoTS_2022}, %our published work in~\cite{Fadul_Access_2021} uses 
five %a set of $N_{C}=20$
``candidate'' preambles are randomly selected from each of $N_{D}=4$ IEEE 802.11a emitters' set of signals to represent the transmitted signal $x(m)$. %Each emitter is represented by five candidate preambles. 
Using the N-M simplex algorithm to solve equation \eqref{eqn:to_be_minimized}--for each candidate preambles--results in a total of $N_{C}=20$ channel impulse response estimates. The residual power formula %presented in~\cite{Kennedy_2010} and 
is given as~\cite{Kennedy_2010},
\begin{equation}
	\hat{h}(m) = \argmin_{c}\left\{\sum_{m}{\big|r(m) - \hat{h}_{c}(m) \ast x_{c}(m)\big|^{2}}\right\},
	\label{eqn:best_estimate}
\end{equation}
and used to select the best channel estimate. Where $\hat{h}_{c}(m)$ is the estimated channel impulse response corresponding to the candidate preamble $x_{c}(m)$, and $1$ $\leq$ $c$ $\leq$ $N_{C}$~\cite{Fadul_WCNC_2019}. 

\subsubsection{MMSE Channel Equalizer%
\label{sec:mmse_equalizer}}
After estimating the channel impulse response, using the N-M estimator, the MMSE algorithm is used to correct for the multipath channel effects. The MMSE equalizer attempts to reconstruct the transmitted signal from the received signal by integrating channel statistics--such as the noise power--along with the estimated multipath channel coefficients \cite{Fadul_InfoSec_2020}. The MMSE's Inclusion of the channel statistics in the equalization process makes it a more robust approach under degrading SNR conditions. The MMSE equalizer aims to reduce the the squared error between the estimated $\hat{x}(m)$ and original $x(m)$ transmitted signals as follows,
\begin{equation}
	\hat{x}(m) = \argmin_{\hat{x}(m)}E\left[(x(m) - \hat{x}(m))^{2}\right].
	\label{eqn:argmin_mmse}
\end{equation} 
If the noise power or channel SNR is known, then the MMSE estimates the transmitted signal by solving,
\begin{equation}
	\hat{\mathbf{x}}_{M} = \mathbf{A}^{H}\left(\mathbf{A} \mathbf{A}^{H} + \gamma^{-1}\mathbf{I}_{A}\right)^{-1}r
	\label{eqn:mmse}
\end{equation} 
where $\mathbf{A}$ is a 2D matrix representing the estimated channel impulse response, $\gamma$ is the SNR, and ${I}_{A}$ is an identity matrix with matching dimensions to $\mathbf{A}$~\cite{Rugini_Letter_2005}.
\subsection{Deep Learning Architectures
\label{sect:DL}}
This section provides brief explanations of each DL architecture and algorithm used to generated the results in Sect.~\ref{sect:Results}.%explains multiple MLP-based DL architectures and algorithms adopted by this work to implement and augment SEI.
\subsubsection{Convolutional Neural Network\label{sect:CNN}}
Convolutional Neural Networks (CNNs) are supervised learning Multi-Layer Perceptron (MLP)-based Neural Networks (NNs) designed to process multi-dimensional data in a grid architecture. It can be used to process a one-dimensional vector such as time-interval data, two-dimensional and three-dimensional grids of pixels such as images~\cite{Goodfellow-2016}. CNNs learn parameters $\theta$ using the back-propagation algorithm to estimate a mapping function of a discriminative model by minimizing a loss function. The loss function computes the error between the model's prediction and the ground truth.

The CNN network is comprised of a MLP prepended with convolutional and pooling layers. Convolutional layers consist of multiple neurons in a grid structure where each neuron corresponds within an %is associated with the result of 
element-wise multiplication of a multi-dimensional kernel (a.k.a., filter) and a region in the input that is the same size as the kernel. The kernel's elements are called weights and they are shared between all neurons~\cite{DL_Practitioners}. In a CNN, convolutional layer(s) are used to detect and extract features from multiple regions of the input data to generate feature maps. Each neuron applies an activation function within the convolutional layer to non-linearly transform the feature map's corresponding element~\cite{DL_Practitioners,Riyaz_CNN_2018}. Pooling layers normally follow convolutional layers to reduce the dimensionality of the activated feature map by computing a statistic summary such as a maximum, minimum, and average of nearby outputs~\cite{Goodfellow-2016}. Max pooling is the most common pooling layer used in CNN networks. In this work, Max Pooling is adopted to extract the maximum-value features within rectangular frames of the activated feature maps. After one or more convolutional and pooling layer stages, fully connected layers (a.k.a., dense layers) are used to detect high level features and pass them onto the output layer~\cite{Restuccia_DeepRadioID_2019}. The purpose of the output layer is to predict the label to which the extracted features belong. In this work CNNs are adopted to implement RF-DNA fingerprint classification using IEEE 802.11a Wi-Fi preambles that transverse a Rayleigh fading channel under degrading SNR.
\subsubsection{Auto-Encoder 
\label{sect:AE}}%
An Auto-Encoder (AE) is an MLP-based NN that attempts to regenerate a multi-dimensional input at the output with as little error as possible~\cite{Masci_CAE,radar}. An AE is an unsupervised learning-based architecture used to learn the distribution and an efficient representation of the input data. Logically, an AE consists of two main parts: the encoder and decoder. Generally, the encoder attempts to estimate a mapping function that outputs an intermediate hidden representation $\mathbf{h} = f(\mathbf{x})$. The decoder aims to reconstruct the input from the hidden representation by applying another mapping function $g(\mathbf{h})$ so the final output is $\mathbf{r} = g(\mathbf{h}(\mathbf{x}))$ is as close as possible to the input $\mathbf{x}$~\cite{Goodfellow-2016}. During training, the AE's loss function penalizes the decoder's output for being different from $\mathbf{x}$. This work uses the Mean Square Error (MSE) as the loss function for all AEs. 
When the encoder is comprised of convolutional and pooling layers, the AE is designated a Convolutional AE (CAE). If the multi-dimensional data (a.k.a., tensor) at the CAE input is $\mathbf{x}$, then the hidden representation corresponding to the $i^{\text{th}}$ tensor is given by,
\begin{equation}
	h_{i} = \beta (x_{i} \ast W + b),
	\label{eqn:CAE_EN}
\end{equation}
where $W$ are the elements (a.k.a., weights) of the convolutional kernel, $b$ is the bias, $\ast$ denotes the convolution operation, and $\beta$ is the activation function~\cite{radar}. The decoder's output $r_{i}$ for $h_{i}$ is given by,
\begin{equation}
	r_{i} = \beta (h_{i} \ast \Tilde{W} + \Tilde{b}),
	\label{eqn:CAE_DE}
\end{equation}
where $\Tilde{W}$ are the deconvolutional kernel weights, and $\Tilde{b}$ is the decoder's bias~\cite{radar}. During training, the CAE parameters--including $W$, $\Tilde{W}$, $b$, and $\Tilde{b}$--are adjusted using backpropagation to minimize the loss function given by,
\begin{equation}
	J(\theta) = \sum\limits_{i=1}^{m}{(x_{i}-r_{i})^2},
	\label{eqn:CAE_MSE}
\end{equation}
where $m$ is the number of training samples.

\subsubsection{Generative Adversarial Networks
\label{sect:GAN}}
A Generative Adversarial Network (GAN) is a DL-based architecture that aims to estimate a generative model by simultaneously training two deep network models using adversarial training. GANs can be used to estimate generative models for multiple applications including: image editing, style transfer, and image synthesis~\cite{GAN_overview_2018}. A GAN is comprised of two models referred to as the Generator, $G$, and the Discriminator, $D$. The $G$ is a generative network that learns the training data distribution so it can generate new samples with the same distribution. The $D$ is a discriminative network tasked with determining whether an input sample belongs to the training data set or was generated by the $G$. GAN training can be viewed as a minimax two-player game in which the $G$ attempts to learn the training data distribution and estimate a mapping function capable of generating new samples that increase the $D$'s probability of making the wrong decision, while $D$ tries to maximize its probability of making the right decision (i.e., differentiating a training sample from a generated one)~\cite{GoodFellow_2014, Goodfellow-2016}. The training results in a unique solution when the $D$'s output is uniformly distributed with a probability of one-half everywhere. 

When MLP networks are used for both the $G$ and $D$ networks, backpropagation can be used to train the entire system where the NN representing the $G$ recovers the data distribution without access to the training data and $D$ maximizes the probability of making the right binary decision~\cite{GoodFellow_2014, GAN_overview_2018}. If the prior probability of $G$ input $\mathbf{z}$ is $P_{\mathbf{z}}(z)$, then the generator mapping function can be given by $G(\mathbf{z};~\theta_{g})$ where $\theta_{g}$ is the MLP parameters of $G$. The $D$ function is given by $D(\mathbf{x};~\theta_{d})$ for input $\mathbf{x}$ and a single output $D(\mathbf{x})$, which represents the probability that $\mathbf{x}$ came from the training data and not the $G$%rather than the $G$ output
~\cite{GoodFellow_2014}. During GAN training, the $G$ network intends to learn the distribution $P_{g}$ over the training data $\mathbf{X}$  by simultaneously maximizing the correct decision probability $D(\mathbf{x})$ and minimizing the term $(1-D(G(\mathbf{z})))$ corresponding to the $G$ network~\cite{GoodFellow_2014}. The GAN minimax optimization problem can be described by the following objective function,
\begin{align}
	\underset{G}{\min} \ \underset{D}{\max}~V(D, G) &= E_{x\sim P_{\text{d}}(x)}\{\log[D(x)]\} \nonumber \\
	&+ E_{z\sim P_{z}(z)}\{\log[1 - D(G(z))]\},
	\label{eqn:GAN_LOSS}
\end{align}
where $E$ is the expected value. The optimum point is reached when the $G$ perfectly recovers the training data distribution (i.e. $P_{g} = P_{data}$)~\cite{GoodFellow_2014, Goodfellow-2016}.
\section{Methodology}
\label{sect:methodology}
This section describes a nontraditional, pre-processing approach that uses %DL semi-supervised learning 
semi-supervised DL to correct for multipath channel effects while simultaneously preserving the SEI exploited RF-DNA fingerprints. %features that are exploited by the RF-DNA fingerprinting-based SEI process. 
In fact, two %This work presents two 
semi-supervised learning-based channel equalization approaches are investigated. %implementation of channel equalization. 
The first approach leverages the GAN architecture's adversarial relationship to train the $G$ %that exists within the GAN architecture to train the $G$ %--using a semi-supervised approach--
such that it learns the distribution of the multipath channel effects as well as the signals' RF-DNA fingerprints to create a %SEI features for the estimation of a 
mapping function capable of estimating and correcting the multipath channel effects. The second approach jointly trains a CAE and CNN architecture--designated herein as JCAECNN--that corrects the %takes advantage of joint CAE and CNN architecture training to correct for the 
multipath channel effects while simultaneously improving the system's ability to extract more discriminative % capability of extracting a more dicriminative 
SEI features.

A Rayleigh fading channel comprised of $L$ paths is applied to each collected IQ preamble to simulate a multipath environment. Each semi-supervised learning approach is used to equalize the resulting multipath preamble set with the goal of maximizing SEI performance. Motivated by the results in~\cite{GlobeCom_Tyler_2021}, semi-supervised learning is conducted using the raw IQ preamble samples as well as their %by augmenting the raw IQ samples with their 
Natural Logarithm (NL) representation. The NL of the raw IQ samples is given by, 
\begin{align}
    \tilde{r}(n) &= \ln[{r_{\mathscr{R}}(n) + jr_{\mathscr{I}}(n)}] = \ln\left[\rho_{r}\exp(j\phi_{r})\right] \nonumber \\
    %&= \ln\left[\rho_{r}\exp(j\theta_{r})\right], \nonumber \\
    &= \ln[\rho_{r}] + j\phi_{r} = \tilde{\rho}_{r} + j\phi_{r},
    \label{eqn:nat_log_calc}
\end{align}
where $r_{\mathscr{R}}$ is the signal's real (In-phase) component, $r_{\mathscr{I}}$ is the imaginary (Quadrature) component, $\rho_{r}$ is the magnitude, and $\phi_{r}$ is the phase angle~\cite{GlobeCom_Tyler_2021}. The $i^{\text{th}}$, equalized preamble's augmented representation--denoted here in as IQ+NL--is,%combined IQ+NL signal representation for the $i^{\text{th}}$ multipath preamble is given by,
\begin{equation}
\resizebox{.85\columnwidth}{!}{$
R^{i}(m,n) = \begin{bmatrix}
r_{\mathscr{R}}^{i}(1,1) & r_{\mathscr{R}}^{i}(1,2) & \cdots & r_{\mathscr{R}}^{i}(1,320) \\
r_{\mathscr{I}}^{i}(2,1) & r_{\mathscr{I}}^{i}(2,2) & \cdots & r_{\mathscr{I}}^{i}(2,320) \\
\tilde{\rho}_{r}^{i}(3,1) &  \tilde{\rho}_{r}^{i}(3,2) & \cdots &  \tilde{\rho}_{r}^{i}(3,320) \\
\phi_{r}^{i}(4,1) & \phi_{r}^{i}(4,1) & \cdots &  \phi_{r}^{i}(4,320)
\end{bmatrix}$},
\label{eqn:preamble_rep}
\end{equation}
\noindent where $i=1,2,\dots,N_{B}$ for a total of $N_{B}$ preambles, % ($N_{B}$),
$m=1,2,\dots,N_{D}$, and $n=1,2,\dots,320$ for a IEEE 802.11a Wi-Fi preamble sampled at %a rate of 
$20$~{MHz}. Combining the preamble's phase behavior--captured by $\phi_{r}$--with the IQ features provides a computationally efficient mechanism that improves SEI performance under degrading SNR~\cite{GlobeCom_Tyler_2021}. The rest of this section provides detailed descriptions fo the %presents the 
two semi-supervised learning approaches that enable SEI using %, pre-processing implementations focused on improving RF-DNA fingerprinting performance using SEI features learned from 
signal's collected under Rayleigh fading and degrading SNR conditions.
\begin{figure*}[!t]
  \centering
  \includegraphics[width=7.0in, height=2.5in]{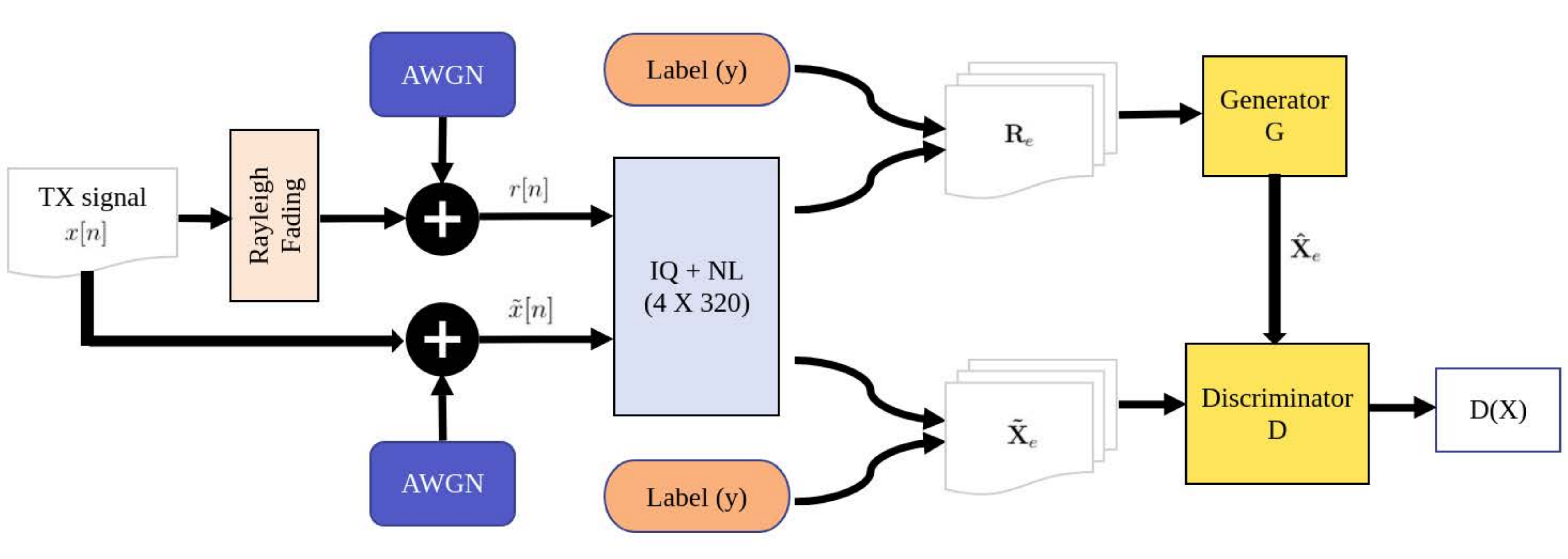}\\
  \caption{\underline{\textit{CGAN Training:}} Flowchart illustrating the process used to train the $G$ to facilitate channel equalization that preserves the exploited, emitter-specific SEI features.}\label{fig:GAN_training}
  \vspace{-5mm}
\end{figure*}
\subsection{Multipath Equalization using a Conditional GAN% 
\label{sec:CGAN_meth}}
Prior to RF-DNA fingerprint generation, all signals undergo channel equalization using a Conditional GAN (CGAN), which is introduced by the authors of~\cite{CGAN_Mizra}. The CGAN is constructed using a CAE and CNN for the $G$ and $D$, respectively. The use of semi-supervised training enables estimation of a generative function $G(z|y)$ capable of reconstructing a Wi-Fi preamble without Rayleigh fading effects despite those effects being present within the preamble input to the $G$. %This implementation uses a Conditional GAN (CGAN) to perform channel equalization prior to RF-DNA fingerprinting~\cite{CGAN_Mizra}. 
Fig.~\ref{fig:GAN_training} shows the CGAN training and RF signal equalization process developed to generate the results presented in Sect.~\ref{sect:Results}. Table~\ref{tab:GAN_arch} provides the configuration and parameters used to construct the CGAN's $G$ and $D$. % architectures of the $G$ and $D$.% shows the overall process developed for CGAN training and RF signal equalization.

Multipath channel effects are induced by filtering each collected preamble using a unique TDL--given in equation~\eqref{eqn:tdl_eqn}--configured to represent a Rayleigh fading channel consisting of $L=5$ reflecting paths. %A Rayleigh fading channel is used to filter each collected preamble $x[n]$ to simulate multipath channel effects. After that, a s
A unique instance of scaled and like-filtered Additive White Gaussian Noise (AWGN) is then added to each preamble to achieve a SNR of 9~{dB} to 30~{dB} in 3~{dB} increments. For each SNR, a total of ten, unique AWGN realizations are generated to augment the data set and facilitate Monte Carlo analysis. The resulting received signal $r[n]$--as expressed in equation~\eqref{eqn:rcvd_signal}--is represented using IQ+NL and each feature is re-scaled to be within the range [0, 1] using Min-Max normalization. The resulting normalized set of $N_{B}$ preambles %--in which each preamble is 4$\times$320 in dimensions--
are then randomly assigned to either the training or test set. The training set is comprised of $N_{R} = 1800$ (i.e., 90\% of the total available at the chosen SNR and noise realization) preambles and the test set consists of $N_{T} = 200$ (i.e., 10\% of the total available at the chosen SNR and noise realization) preambles where $N_{T} = N_{B} - N_{R}$.% A total of $N_{B_{R}}$ are randomly selected from the $N_{B}$ preambles to train the CGAN and RF-DNA fingerprinting classifier. The remaining $N_{B_{T}} = N_{B} - N_{B_{R}}$ preambles not selected for training are use to blindly test the model performance.     

The CGAN is first proposed in \cite{CGAN_Mizra} and is an extension of the traditional GAN introduced in \cite{GoodFellow_2014}. In CGAN, the $G$ and $D$ networks--shown in Fig.~\ref{fig:GAN_training}--accept the class label $y$ as an additional input; % $y$ as the class label. 
thus, both mapping functions $G(z)$ and $D(X)$ are conditioned on the variable $y$ and the traditional GAN's minimax  
optimization equation is rewritten as \cite{CGAN_Mizra},
\begin{align}
	\underset{G}{\min} \ \underset{D}{\max}~V(D, G) &= E_{x\sim P_{\text{d}}(x)}\{\log[D(x|y)]\} \nonumber \\
	&+ E_{z\sim P_{z}(z)}\{\log[1 - D(G(z|y))]\},
	\label{eqn:CGAN_LOSS}
\end{align}
where $P_{\text{d}}(x)$ is the training data distribution learned by the $G$. In this work, %the variable $y$ represents 
the class label %and 
is combined with the input to the $G$ and $D$ using a hidden representation that enables the GAN to estimate a one-to-many generative function conditioned by $y$ instead of the traditional one-to-one mapping. The hidden representation is generated by an embedding layer that maps each emitter's class label to a length fifty vector. It is important to note that the length of the vector is empirically chosen based on \cite{Gereme_embedding_2020}. The label vector size is %then 
expanded to a length of 1,280 using a dense layer and reshaped into a 4$\times$320 tensor to match the dimensionality of its assigned preamble. %This two-dimensional 
The 4$\times$320 label is appended to the corresponding emitter's normalized IQ+NL preamble representations to form a 4$\times$320$\times$2 labeled preamble representation denoted as $\mathbf{R}_{e}$.

\begin{table}[!b]
\vspace{-5mm}
\centering
\caption{The configuration for the neural networks associated with the Generator $G$, the discriminator $D$ and RF-DNA fingerprint classifier used for CGAN equalized RF-DNA Fingerprinting.}
\includegraphics[width=0.9\columnwidth]{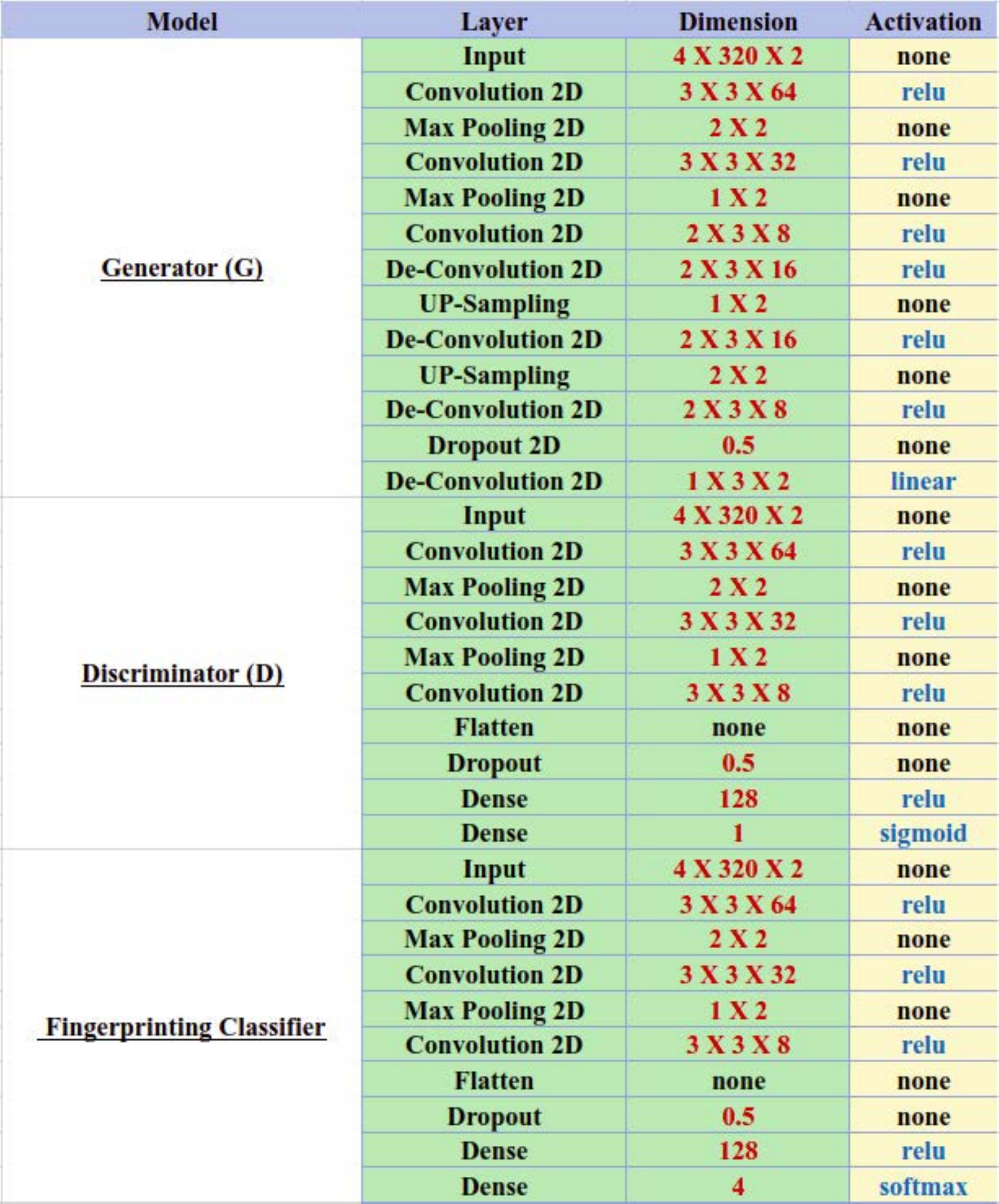}
% \vspace{-5mm}
\label{tab:GAN_arch}
\end{table}

\begin{figure}[!t]
  \centering
%   \vspace{-5mm}
  \includegraphics[width=\columnwidth]{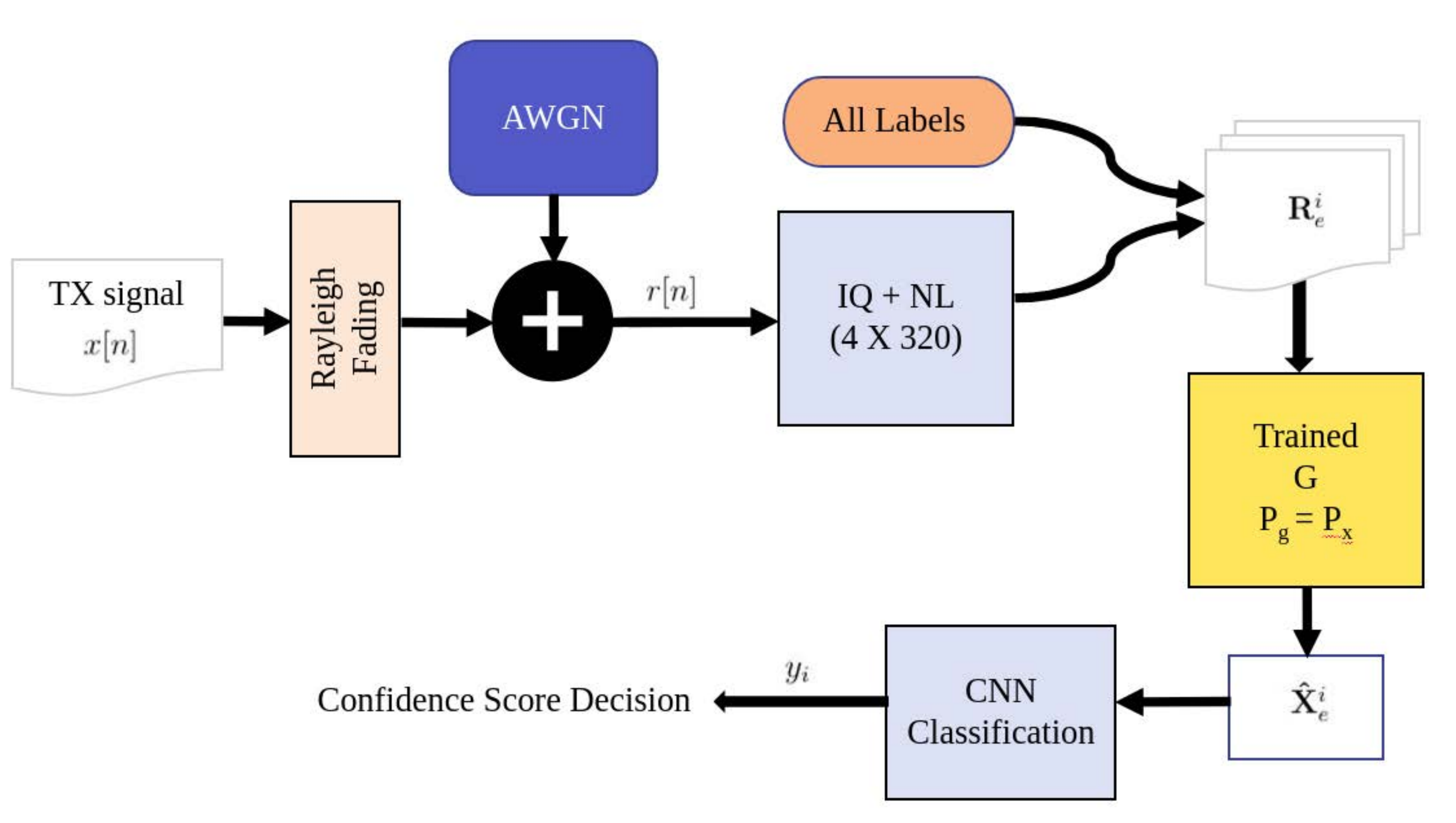}
  \caption{Illustration of the SEI process that uses a trained CGAN's $G$ network to perform multipath channel equalization while simultaneously preserving emitter specific RF-DNA fingerprinting exploited features prior to CNN classification.
  \label{fig:GAN_fingerprinting}}
  \vspace{-5mm}
\end{figure}

During CGAN training, the $G$'s input is the preamble $\mathbf{R_{e}}$ %input to the are applied at the input of the $G$ 
and the $D$'s input is a set of \textit{AWGN-only} preambles $\mathbf{\tilde{X}}_{e}$ (i.e., there are no multipath effects present within this signal set). The $G$ attempts to learn the multipath impacted preambles' distribution and estimate a function $G(z)$ that maps them to $\mathbf{\hat{X}}_{e}$ such that the distribution of $\mathbf{\hat{X}}_{e}$ matches the distribution associated with the \textit{AWGN-only} preambles $\mathbf{\tilde{X}}_{e}$ used to train the $D$.
The $D$ outputs a single-value $D(x)$ representing the probability that the input $x$ is from the training data $\mathbf{\tilde{X}}_{e}$ rather than %the $G$ output 
$\mathbf{\hat{X}}_{e}$. The CGAN %in Fig.~\ref{fig:GAN_training} 
is trained using backpropagation with a minibatch  size of 256 tensors, 10,000 epochs, and an alternating scheme in which the $D$ is trained for $S$ steps based upon a given $G$. % for $S$ steps. 
Training for $S$ steps results in the best version of the $D$. %for a given $G$. 
The authors of~\cite{GoodFellow_2014} treat $S$ as a hyperparameter with $S=1$ being the least computationally complex; thus, $S$ is set to one for the results presented in Sect.~\ref{sect:Results}. The $D$ is trained using forward- and backpropagation with the goal of maximizing $V(D, G)$ to achieve the highest correct decision probability. After $S = 1$ steps. The $G$ is trained using Stochastic Gradient Descent (SGD) to minimize $V(D, G)$ with the goal of reducing the $D$'s ability to make a correct decision. This training process continues until $D(x)=0.5$ everywhere or the total number of training iterations equals the empirically chosen value of 10,000. \\
\indent Once the CGAN is trained, the resulting $G$--that provides the mapping function $G(z|y, \theta_{g})$--is disconnected from the $D$ and used to equalize the multipath preambles $r[n]$ generated from the $N_{T}$ preambles comprising the test set of $x[n]$. Each test preamble $r[n]$ is combined with each of the $N_{D}=4$ labels using the hidden representation--described earlier in this section--to create a total of four labeled preambles $\mathbf{R}^i_{e}$ where $i=1,2,3,4$. The trained $G$ generates an equalized output $\mathbf{\hat{X}}^{i}_{e}$ corresponding to each labeled preamble $\mathbf{R}^i_{e}$. Finally, the second channel (a.k.a., the hidden representation label) is removed from the $G$'s outputs $\mathbf{\hat{X}}^{i}_{e}$ prior to classification using a trained CNN. This process is illustrated in Fig.~\ref{fig:GAN_fingerprinting}. \\
\indent The CNN is trained using: (i) $4\times320$ tensors formed using the IQ+NL preamble representations generated from the \textit{AWGN-only} training set $\tilde{x}[n]$, (ii) backpropagation, (iii) SGD to minimize the categorical, cross-entropy loss function, (iv) $l_{2}$-norm regularization to reduce overfitting, and (v) Adam optimization for the adjustment of the network's weights~\cite{adam_optimizer}. The CNN's output layer uses a softmax decision to assign the $i^{\text{th}}$ equalized representation $\mathbf{\hat{X}}^{i}_{e}$ a label of $y_{i}$ according to,
\begin{equation}
	y_{i} = \max_{j}(Q_{ij}),
	\label{eqn:softmax}
\end{equation}
where $Q_{ij}$ is the $j^{\text{th}}$ CNN output for the $i^{\text{th}}$ equalized representation input $\mathbf{\hat{X}}^{i}_{e}$, $j=[1,2, 3, 4]$ is the index of the softmax layer outputs, and $i=[1,2, 3, 4]$ for a given received preamble $r[n]$. Lastly, a confidence score decision is used to assign the received multipath preamble $r[n]$ a label $\hat{y}$ that satisfies,
\begin{equation}
	\hat{y} = \max_{i}( y_{i} ) = \max_{i}\max_{j}(Q_{ij}).
	\label{eqn:fprint}
\end{equation}
%
%\noindent where $Q_{ij}$ is the $j^{\text{th}}$ CNN output for the $i^{\text{th}}$ equalized representation input $\mathbf{\hat{X}}^{i}_{e}$. 
%
\begin{figure*}[!t]
  \centering
  \includegraphics[width=5.0
  in]{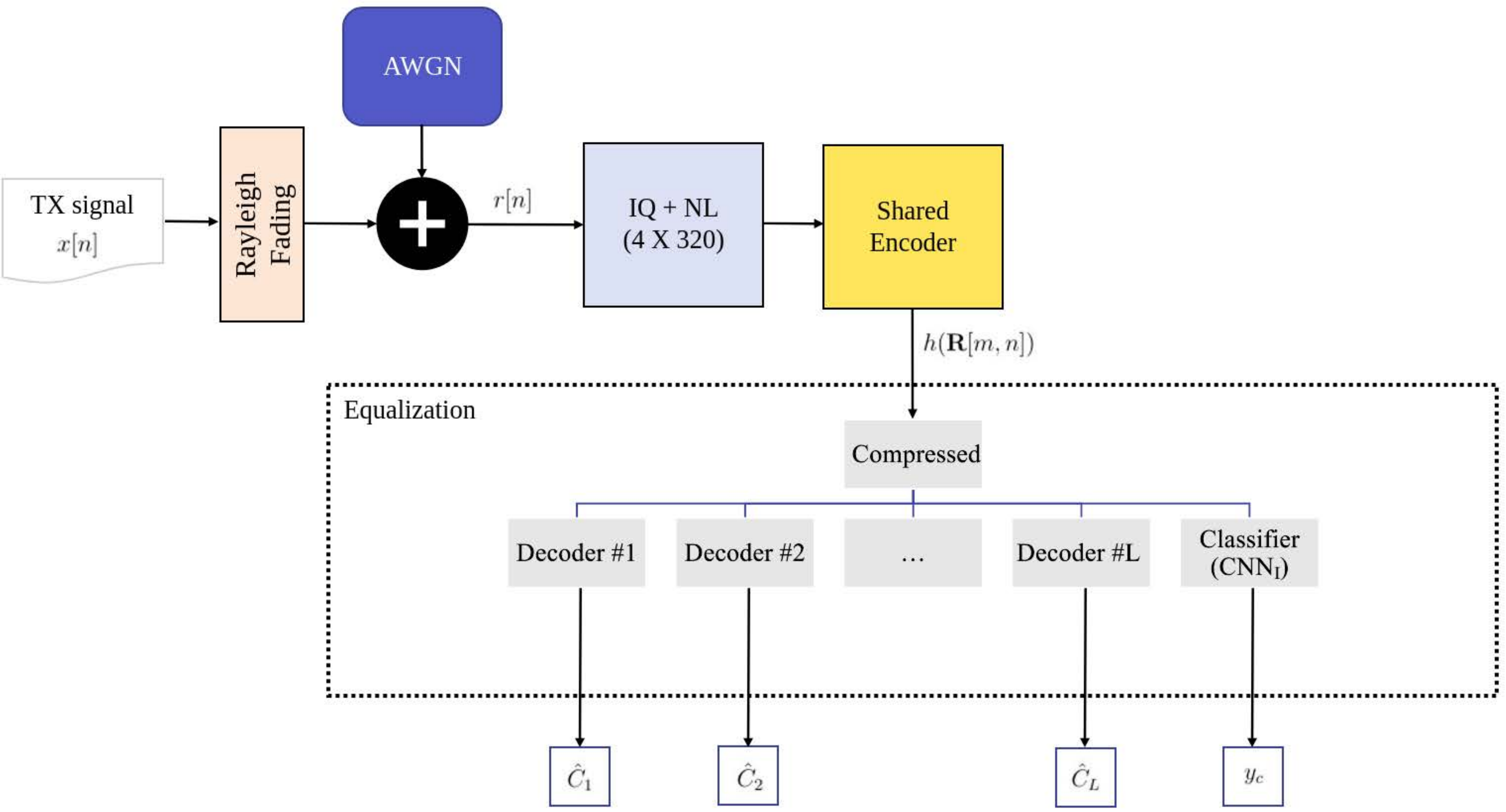}
  \caption{\underline{\textit{JCAECNN Training:}} Flowchart illustrating signal pre-processing and the joint CAE multipath equalization.
  \label{fig:Custom_Loss_training}}
  \vspace{-5mm}
\end{figure*}
\indent The best possible SEI performance is achieved by training the CNN at SNRs lower than those in the test set. A grid search is performed to determine the best SNR at which to train the CNN and achieve the highest SEI performance across all SNR values. For example, the CNN is trained using preambles collected at an SNR of 12~{dB} and once trained that CNN classifies preambles collected at each SNR in the range from 9~{dB} to 30~{dB} with 3~{dB} increments between consecutive SNR values. The `best' training SNR is the SNR whose corresponding trained CNN results in the highest average percent correct classification across all SNR values. The CGAN SEI process can be summarized using the following algorithm,
%
\begin{comment}
\begin{algorithm}
\caption{SEI using CGAN Equalization}
\begin{algorithmic}[1]

\Procedure {CGAN Training}{$x[n]$, $N_{B}$, $y$}       
    \State Create Dataset1 $r[n]$: 
        \State Extract x[n]
        \State Convolution with $L=5$ Rayleigh TDL
        \State Add AWGN for SNR $\in$ [9, 30]~dB \& $N_{z}=10$
    \State Create Dataset2 $\tilde{x}[n]$:
        \State Extract x[n] 
        \State Add AWGN for SNR $\in$ [9, 30]~dB \& $N_{z}=10$      
    \State $\mathbf{R}_{e}$ = IQ+NL($r[n]$) + embedding(y)          \Comment{Create Labeled preambles from r[n]}
    \State $\mathbf{\tilde{X}}_{e}$ = IQ+NL($\tilde{x}[n]$) + embedding(y)          \Comment{Create Labeled preambles from r[n]}
    \State Split Datasets: $N_{R} = 0.9 \times N_{B}$ \& $N_{T} = N_{B} - N_{R}$   \Comment{Training \& Test sets}
    \For{\texttt{<SNR $\in$ [9, 30]~{dB}>}}
        \For{\texttt{<epoch = 1 ; epoch <= 10,000>}}
        \State $\mathbf{\hat{X}}_{e}$ = $G(\mathbf{{X}}_{R_{e}})$
        \For{\texttt{<Steps = 1 ; Steps <= $S$>}}
        \State D(x) = $D(\mathbf{\hat{X}}_{e}) ~\|~D(\mathbf{\tilde{X}}_{e})$
        \State Update: $D$ Trainable Parameters
        \EndFor
        \State Update: $G$ Trainable parameters based on $D$ output
      \EndFor
      \State Save Trained $G$
    \EndFor
\EndProcedure

\Procedure {CNN Training}{Trained $G$, $N_{B}$, $y$}

\EndProcedure
\end{algorithmic}
\end{algorithm}
\end{comment}
%
\subsection{Multipath Equalization using JCAECNN% 
\label{sec:JCAECNN_meth}}
This equalization approach uses a joint CAE and CNN (a.k.a., JCAECNN) architecture similar to that used in~\cite{Yu_WiMob_2019}; however, the approach described here differs from that in~\cite{Yu_WiMob_2019} in that the CAE architecture is modified to consist of multiple decoder heads instead of one. The decoder heads %facilitate the decomposition of a
decompose a received preamble--that undergoes Rayleigh fading--into its $L$ weighted and delayed copies of the transmitted signal, $x[n]$. As illustrated in Fig.~\ref{fig:Custom_Loss_training}, the JCAECNN equalization process is built on a Single Input Multiple Output (SIMO) system that includes: signal collection and detection (Sect.~\ref{sect:Signal_of_Interest}), multipath fading and AWGN scaling~(Sect.~\ref{sect:Multipath_Model}), and signal preparation--in accordance with our prior work~\cite{GlobeCom_Tyler_2021}--prior to equalization and classification.

%\begin{enumerate}%[leftmargin=*]
% \item{\textit{Multipath Fading:} Convolution of the collected signals $x[n]$ with a Rayleigh fading channel comprised of $L$ paths to simulate multipath channel effects.} 
% \item{\textit{Noise Scaling:} Like-filtered and scaled white Gaussian noise is added to the collected signals to achieve the selected SNR value. This process is repeated to achieve SNR values from 9~{dB} to 30~{dB} in increments of 3~{dB} between consecutive SNR values. For each SNR, a total of ten noise realizations are generated to facilitate Monte Carlo simulation and assessment, as well as augmenting the RF signal data set.}
% \item{\textit{Signal Preparation:} For each SNR and every noise realization, the resulting noisy multipath preamble $r[n]$ is extracted and the NL is calculated using equation \eqref{eqn:nat_log_calc} and combined with the IQ samples to form the $4\times320$ IQ+NL signal representation $R[m,n]$ as given by \eqref{eqn:preamble_rep}. Min-Max normalization is then used to scale each feature within each preamble to the range [0, 1]. A total of $N_{B_{R}}$ are randomly selected from the $N_{B} = 2000$ preambles to train the CAE and CNN$_{I}$ in Fig.~\ref{fig:Custom_Loss_training}. The remaining $N_{B_{T}}$ preambles not selected for training are used from blind testing}. 
%\item{
Equalization is performed using a CAE consisting of a single encoder and $L$ decoder heads (i.e., one for each path of the Rayleigh fading channel) using the preambles' %performs multipath equalization on the 
IQ+NL representations %of the collected set of preambles that exhibit Rayleigh fading effects 
at the selected SNR. Motivated by the results presented in~\cite{DenseNet}, a Densely Connected Convolutional Network (DenseNet) is used to implement the ``shared'' encoder shown in Fig.~\ref{fig:Custom_Loss_training}. DenseNet is comprised of multiple Dense blocks where each block is created by connecting each convolutional layer to all preceding convolutional layers of the same size.  
 These dense connections grant each convolutional layer access to all previously generated feature maps to enable feature reuse. In this work, two densely connected blocks are used to construct %represent 
the encoder. The DenseNet encoder generates compressed representation $h(\mathbf{R}[m,n])$, which is the input shared with each of the following NN architectures: the $L$ decoder heads and the classifier %\todo{What is with the subscript I? What does it mean?}\hl{CNN$_{I}$}. %represents an input between all the following NN architectures including a total of $L$ decoder heads and the CNN$_{I}$ classifier.
The resulting JCAECNN architecture is derived from the square error function of %in 
equation~\eqref{eqn:to_be_minimized}. The optimization goal is to minimize the error between the received signal $r(m)$ and its delayed and scaled copies $C_{k} = \alpha_{k} x(m - \tau_{k})$ %where $k = 1, 2, \cdots, L$ 
generated using the TDL of equation~\eqref{eqn:tdl_eqn}. %a length $L$ Rayleigh fading channel. 
The target of each decoder head is to reconstruct a single copy $C_{k}$ corresponding to the $k^{\text{th}}$ path of the Rayleigh fading channel. In addition to the decoder heads, a CNN classifier (CNN$_{I}$) is used to assign the compressed representation $h(\mathbf{R}[m,n])$ to any of the $N_{D}=4$ emitters using a softmax decision. The NN configurations for the shared encoder, decoder heads, and CNN$_{I}$ classifier are provided in Table~\ref{tab:Custom_Loss_arch}. \\ 
%}
%\end{enumerate}
%
\begin{table}[!b]
\vspace{-5mm}
\centering
\caption{The configuration for the neural networks associated with the shared DenseNet Encoder, Decoder heads, and RF-DNA fingerprint classifier for the JCAECNN RF-DNA Fingerprinting}
\includegraphics[width=0.95\columnwidth]{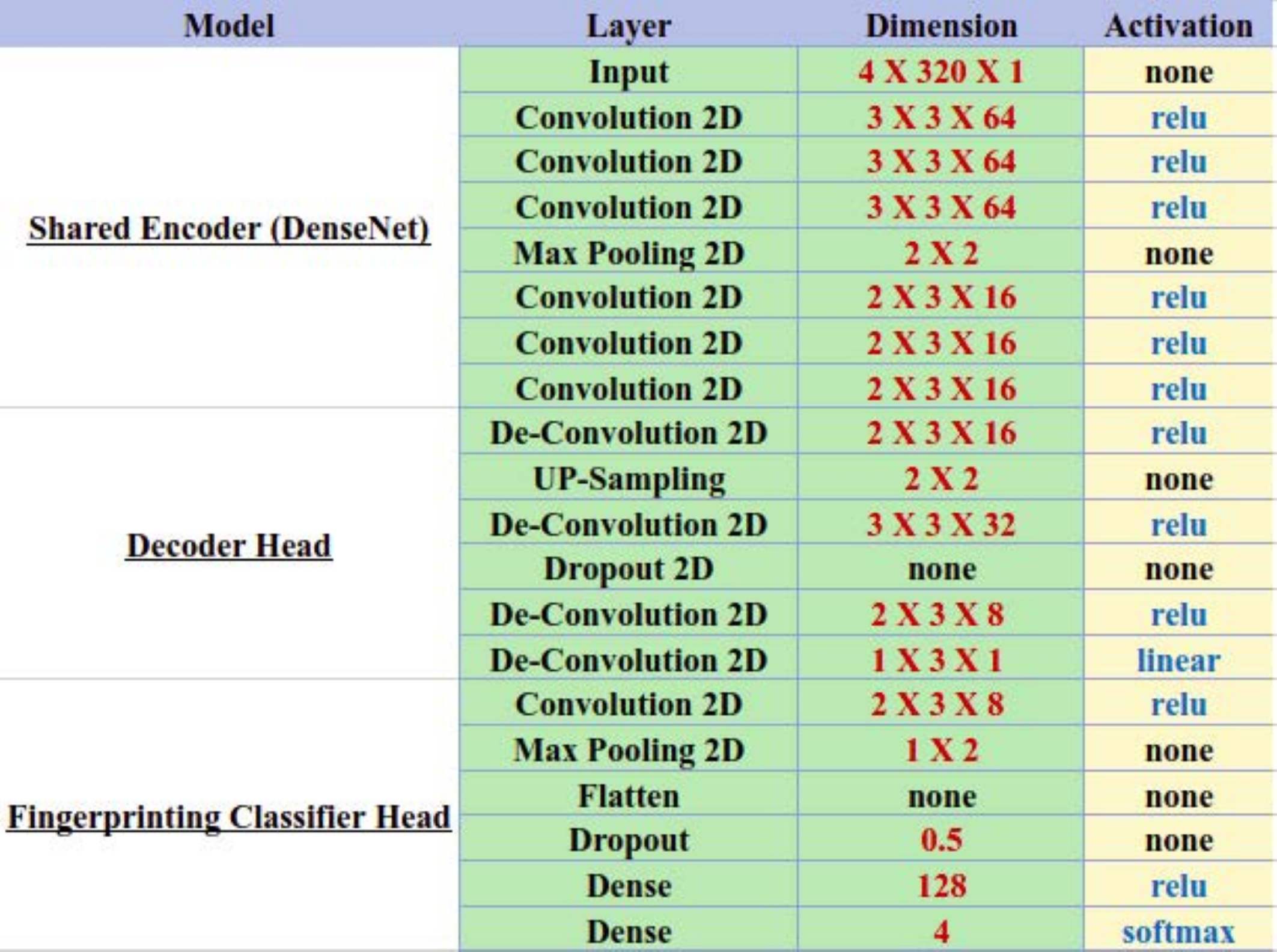}
% \vspace{-5mm}
\label{tab:Custom_Loss_arch}
\end{table}
\indent The %Single Input Multiple Output (SIMO) system in~Fig.~\ref{fig:Custom_Loss_training} 
JCAECNN is trained iteratively to jointly optimize the %by optimizing a 
combined loss function given by,
%
% \begin{align}
% 	F(\theta) &= \sum_{k=1}^{L} (\lambda_{k} \times M(x(m - \tau_{k}), \hat{x}(m - \tau_{k}))) 
% 	\nonumber \\
% 	&\qquad{} + \lambda_{c} \times C(y_{c}, \tilde{y_{c}})
% 	\label{eqn:custom_loss}
% \end{align}
%
\begin{equation}
    F(\theta) = \sum\limits_{k=1}^{L} \left[\lambda_{k} \times M(m,\tau_{k})%x(m - \tau_{k}), \hat{x}(m - \tau_{k}))) 
	+ \lambda_{c} \times C(y_{c}, \tilde{y_{c}})\right],
	\label{eqn:custom_loss}
\end{equation}
where $\lambda_{k}$ are the $k^{\text{th}}$ decoder head's loss weights corresponding to each Rayleigh fading path, $\lambda_{c}$ are CNN's loss weights, $M(m,\tau_{k})$ is the MSE loss given by,
\begin{equation}
M(m,\tau_{k}) = \dfrac{1}{N_{s}} \sum\limits_{m=1}^{N_{s}} (x(m-\tau_{k}) - \hat{x}(m-\tau_{k}))^{2}	
\end{equation}
where $\hat{x}(m - \tau_{k}) = \hat{C}_{k}$ is the decoder's output corresponding to the $k^{\text{th}}$ path, and $C(y_{c}, \hat{y}_{c})$ is the categorical cross entropy function used by the CNN classifier to compute the difference between the ground-truth label $y_{c}$ and estimated label $\hat{y}_{c}$. If the ground-truth label is one-hot encoded where $y_{c}=[y_{c,1}, y_{c,2}, \dots, y_{c,N_{D}}]$, the categorical cross entropy can be calculated by the following formula, 
\begin{equation}
C(y_{c}, \hat{y}_{c}) =  -\sum\limits_{l=1}^{N_{D}} (y_{c,l} \times \log (\hat{y}_{c,l})) 
\end{equation}
where $l$ is the index of the class at both the one-hot encoded ground truth and the output layer of the classifier \cite{Polat_CCE_2022}. %Joint training is achieved by iterating across each decoder head and the CNN classifier. 
An individual fading path's MSE loss is optimized by updating the weights of the shared encoder and decoder head assigned to the chosen fading path.\\%The SIMO system is trained in an alternating scheme where each individual path's loss is optimized to update the weights of the shared Encoder and the corresponding Decoder head.
\indent Decoder head training allows the shared encoder to learn compressed and delay-invariant SEI features. %Training the SIMO system to reconstruct the copies $C_{k}$ by the corresponding decoder heads from the shared compressed representation $h(\mathbf{R}[m,n])$ enables the the encoder to learn SEI features that are delay-invariant. 
The CNN classifier head %in the SIMO system 
is trained to minimize %optimize 
the $C$ loss between the actual label $y_{c}$ and the estimated label $\tilde{y_{c}}$, which allows %CNN$_{I}$ training enables 
the shared encoder to learn  more discriminating SEI features than those %ones 
learned while training the decoder heads.\\
%due to decoder heads optimization. The JCAECNN training process %Therefore, the proposed joint training of the SIMO architecture in Fig.~\ref{fig:Custom_Loss_training} mitigates Rayleigh fading channel effects by forcing the shared encoder to simultaneously learn: (i) a compressed and delay-invariant signal representation that is used by the $L$ decoder heads to generate an equalized version of the received signal \todo{Correct for notation as needed.}\hl{$r(t)$}, and %of the signal that facilitates the reconstruction (channel correction) of the transmitted signal using the decoder heads, and (ii) enabling the shared encoder to learn (ii) more discriminative features along with time-invariant features improve the over all RF-DNA fingerprinting performance when the RF signals undergo Rayleigh fading channel.
%
\begin{figure}[!b]
  \centering
  \vspace{-5mm}
  \includegraphics[width=\columnwidth]{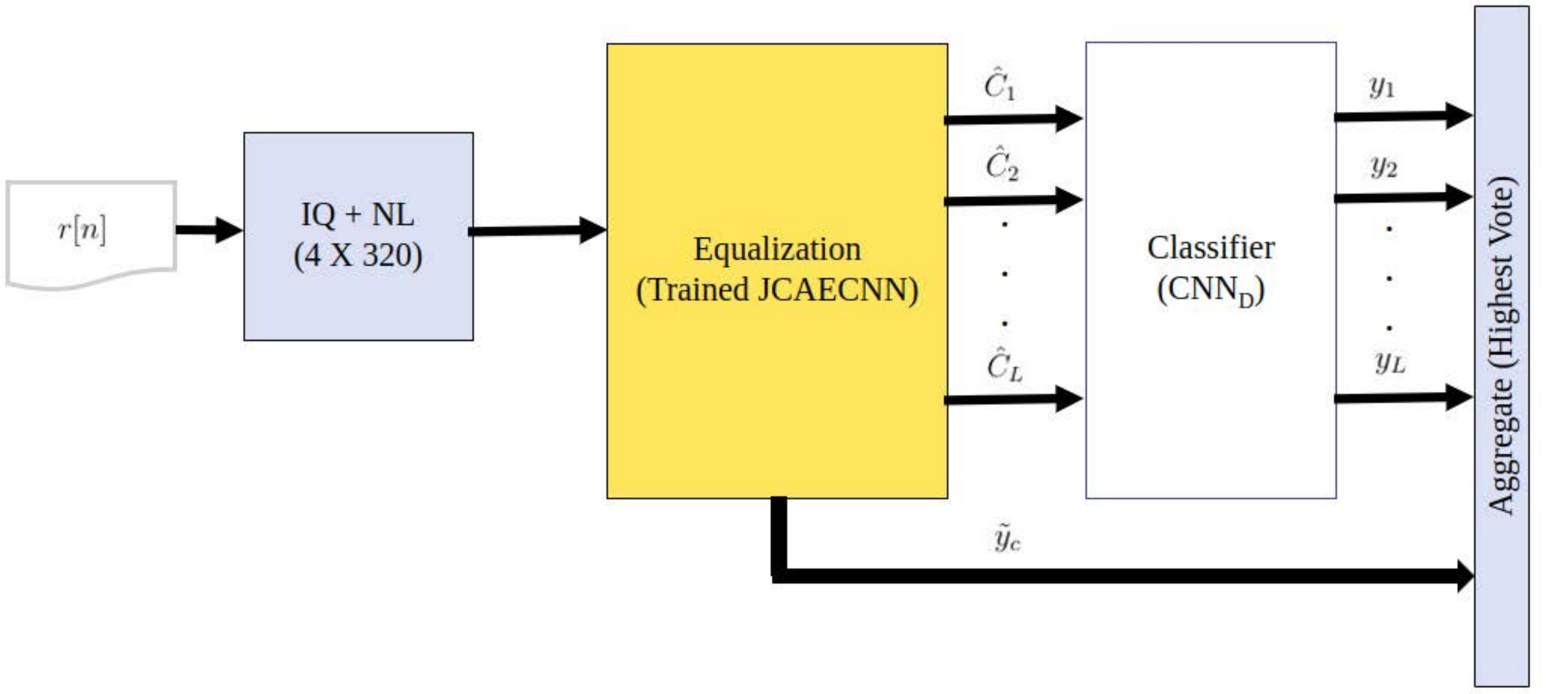}
  \caption{\underline{\textit{JCAECNN Testing:}} Flowchart illustrating the RF-DNA fingerprinting process using the trained JCAECNN system that performs channel equalization and supplies the CNN$_{I}$ decision for an aggregated CNN classification decision.} 
  \label{fig:Custom_Loss_Fprnt}
\end{figure}
\indent As shown in Fig.~\ref{fig:Custom_Loss_Fprnt}, the trained JCAECNN supplies the equalized preambles $\hat{C}_{k}$--for $k = 1, 2, \cdots, L$--along with the CNN$_{I}$ decision $\tilde{y_{c}}$ for every normalized IQ+NL preamble representation $\mathbf{R}[m,n]$ to the discriminating CNN, which is denoted as CNN$_{D}$. CNN$_{D}$ assigns $\hat{C}_{k}$ label $y_{k}$ for all $k$ and decides that the received preamble was transmitted by the emitter whose class label achieves the highest vote out of $y_{k}$ and $y_{c}$. The CNN$_{D}$ architecture is similar to the RF-DNA fingerprint classifier architecture shown in Table~\ref{tab:GAN_arch} and trained using the same data set used to train the JCAECNN with only AWGN present (i.e., there are no multipath effects present within the signal set).

\begin{comment}
After training, the SIMO system in Fig.~\ref{fig:Custom_Loss_training} is used to preprocess the normalized IQ+NL preambles $\mathbf{R}[m,n]$ of the test set prior to RF-DNA fingerprinting as shown in Fig.~\ref{fig:Custom_Loss_Fprnt}. For each preamble representation $\mathbf{R}[m,n]$, the trained SIMO system generates the following outputs:
%
\begin{enumerate}%[leftmargin=*]
\item{Estimate of the scaled and delayed signals $\hat{C}_{k}$ where $k = 1, 2, \cdots, L$ for a length $L$ Rayleigh fading channel. Each signal estimate $\hat{C}_{k}$ is generated by the corresponding decoder head.}
\item{Estimate label $\tilde{y}_{c}$ of the transmitted preamble $x[n]$. $\tilde{y}_{c}$ is generated by the CNN$_{I}$ classifier.}
\end{enumerate}

A CNN classifier denoted by CNN$_{D}$ is used to assign each of the signal estimates $\hat{C}_{k}$ to a label $y_{k}$. Finally, the received preamble $r[n]$ from the blind test set is assigned to the label that achieves the highest vote out of $y_{k}$ and $y_{c}$. 
The CNN$_{D}$ NN architecture is similar to the RF-DNA fingerprint classifier in Table~\ref{tab:GAN_arch}. At each SNR, CNN$_{D}$ is trained using IQ+NL representation of a data set comprised of ($x[n], x[n-\tau_{1}], \cdots, x[n-\tau_{L}]$) corresponding to the same $N_{B_{R}}$ preambles used to train the SIMO system in Fig.~\ref{fig:Custom_Loss_training}. 
\end{comment}

\section{Results%
\label{sect:Results}}%
This section presents the experimental results and analysis for the RF-DNA fingerprint-based SEI of IEEE 802.11a Wi-Fi emitters whose signals transverse a Rayleigh fading channel and degrading SNR using the CGAN and JCAECNN processes described in Sect.~\ref{sec:CGAN_meth} and Sect.~\ref{sec:JCAECNN_meth}, respectively. Results for the following experiments are presented.
\begin{enumerate}[leftmargin=*]
    \item{\underline{\textit{Experiment \#1:}} Comparative assessment is conducted between the CGAN and JCAECNN approaches and our RF-DNA fingerprinting of emitters under Rayleigh fading published in~\cite{Fadul_Access_2021,Fadul_GIoTS_2022}.}
    \item{\underline{\textit{Experiment \#2:}} Scalability analysis of the CGAN and JCAECNN approaches. This experiments assesses SEI performance for both approaches using data sets consisting of signals collected from: four, eight, sixteen, and thirty-two commercial emitters to represent larger IoT device deployments.}
    \item{\underline{\textit{Experiment \#3:}} Assessment of JCAECNN SEI performance using the sixteen emitter signal set of Experiment~{\#2} as well as the publicly available data sets associated with the published results presented in~\cite{Oracle_2019,WiSig}. This experiment %was conducted to 
    evaluates the JCAECNN architecture's effectiveness in learning SEI discriminating features from multipath signals whose transmitting emitters differ in manufacturer, model, or Size, Weight, and Power-Cost (SWaP-C) with respect the to the emitters described in Sect.~\ref{sec:data_collection}. %the performance of the proposed SEI approach to discriminate between different types of emitters located on different environments. 
    Furthermore, the results of this experiment serve as a %type of analysis provides a 
    benchmark to facilitate the evaluation of future work.}
    \item{\underline{\textit{Experiment \#4:}} This experiment assesses JCAECNN SEI performance using a loss function whose weights are selected based upon the Rayleigh fading channel's path variances. This experiment optimizes the MSE loss weights to maximize JCAECNN performance.%by optimizing the MSE loss weights.
    }
\end{enumerate}
The specifics of the data set(s), training, and testing approaches are explained within each experiment's section.
\subsection{Results for Experiment \#1}
\label{sec:results_exp01}
The IEEE 802.11a Wi-Fi signal set used in this experiment is comprised of preambles extracted from the four Cisco AIR-CB21G-A-K9 Wi-Fi emitters described in Sect.~\ref{sec:data_collection} and used to generate the results presented in our prior publications~\cite{Fadul_Access_2021,Fadul_GIoTS_2022}. A total of 2,000 preambles are extracted for each of the four emitters, which is subdivided into a training and testing set comprised of $N_{R} = 1,800$ and $N_{T} = 200$ randomly selected preambles per emitter. The training set is duplicated to permit creation of a \textit{AWGN-only} set--that is used in the training of the CGAN--and a Rayleigh fading plus AWGN set. The test set is comprised of only Rayleigh fading plus AWGN impacted preambles. It is important to note that the Rayleigh fading channel consists of $L = 5$ paths, is generated and applied to each signal as described in Sect.~\ref{sect:Multipath_Model}, and is unique to each preamble (i.e., the channel coefficients change every transmission). A specific SNR is achieved by adding scaled and like-filtered AWGN to every preamble and the process repeated ten times to permit Monte Carlo simulation. SEI performance is assessed using average percent correct classification performance at SNR values ranging from 9~{dB} to 30~{dB} in steps of 3~{dB} between consecutive values. 

Training of the CGAN and JCAECNN approaches is conducted in accordance with Sect.~\ref{sec:CGAN_meth} and Sect.~\ref{sec:JCAECNN_meth}, respectively. Recall that CGAN SEI performance is maximized by training it using SNR values lower than those comprising the test set. The results of the grid search--described in Sect.~\ref{sec:CGAN_meth}--are presented in Table.~\ref{tab:SNR} in which SNR$_{R}$ designates the SNR of the preambles used to train the CGAN and SNR$_{T}$ is the SNR of the preambles being classified by the CGAN trained at SNR$_{R}$.
\begin{table}[!t]
% \vspace{-5mm}
\centering
\caption{The training SNR, dentoed as SNR$_{R}$, that maximizes the CGAN SEI performance when classifying test set preambles at %corresponding to each classification SNR (SNR$_{T}$)
SNR$_{T}$. All SNR values are in decibels.}
\includegraphics[width=0.9\columnwidth]{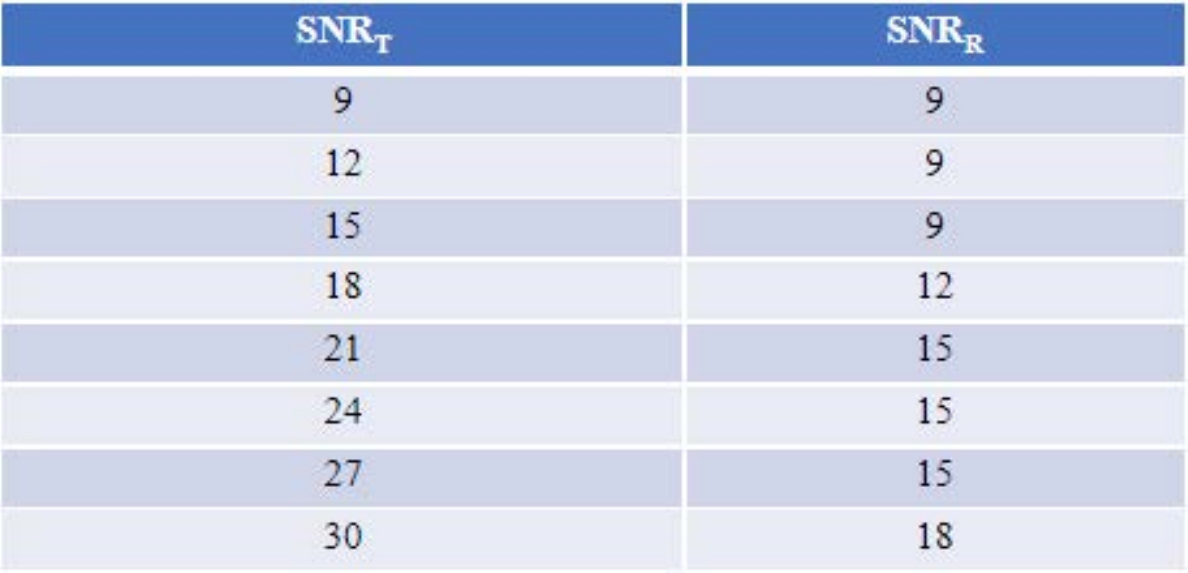}
\vspace{-5mm}
\label{tab:SNR}
\end{table}
The JCAECNN approach's ``best'' training SNR is determined using the same approach as that of the CGAN and resulted in a SNR$_{R}$ value of 9~{dB} for all SNR$_{T}$ values. 
The Partitioned Time RF Fingerprinting (PTRFF) approach from~\cite{Fadul_Access_2021} and traditional Time-Frequency Feature-Engineered RF Fingerprinting (TFRFF) approach from~\cite{Fadul_GIoTS_2022} both perform channel correction using the N-M channel estimator and MMSE channel equalizer described in Sect.~\ref{sec:nm_estimator} and Sect.~\ref{sec:mmse_equalizer}, respectively. The PTRFF approach uses a pre-trained one-dimensional CNN to classify partitioned, time IQ preambles. The time partitions are generated by slicing the IQ samples of each equalized preamble using a $N_{w} = 64$ length, sliding window. % of length $N_{w} = 64$. 
The TFRFF approach performs feature extraction from the normalized, Gabor coefficients calculated from a preamble's IQ samples and classification performed using the Multiple Discriminant Analysis/Maximum Likelihood (MDA/ML) classifier. The reader is directed to \cite{Fadul_Access_2021} and \cite{Fadul_GIoTS_2022} for the specific methodologies, results, analysis, and conclusions of the PTRFF and TFRFF approaches, respectively. \\
\begin{figure}[!t]
	\centering
% 	\vspace{-5mm}
	\includegraphics[width=0.45\textwidth]{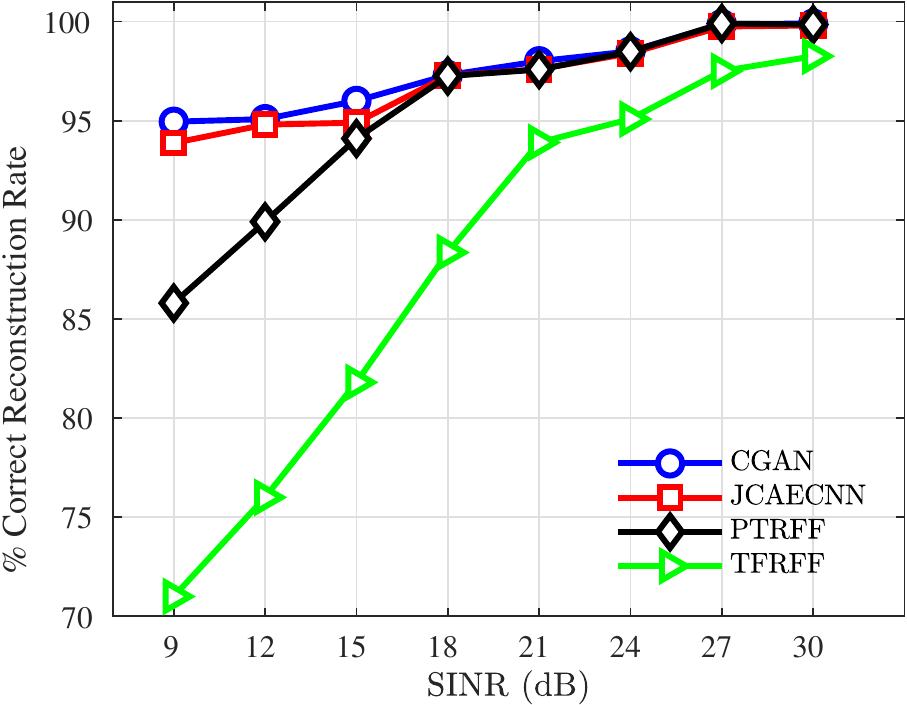}
	\caption{\underline{\textit{Experiment \#1 Results:}} Average percent correct classification performance across the $N_{D}=4$ IEEE 802.11a Wi-Fi emitters using the CGAN, JCAECNN, PTRFF, and TFRFF approaches %time partitioned, $T_{P}$, preambles with a sliding window of length $N_{w}=64$; and feature-engineered $TF_{FE}$ 
	for SNR$\in$[9, 30]~{dB} in steps of 3~{dB}.}
	\label{fig:DL_Trad_Comp}
	%\medskip
	\vspace{-5mm}
\end{figure}
\indent Average percent correct classification performance for the CGAN, JCAECNN, PTRFF, and TFRFF SEI approaches are presented in Fig.~\ref{fig:DL_Trad_Comp}. Each data point is associated with $N_{T} \times N_{z} \times N_{D} = 8,000$ individual classification decisions. The results of the JCAECNN are generated by setting $\lambda_{k}$ and $\lambda_{c}$ to one.
Compared to the PTRFF, and TFRFF approaches, the results in %It can be seen from 
Fig.~\ref{fig:DL_Trad_Comp} show that the %both CGAN- and JCAECNN-based 
CGAN and JCAECNN approaches achieve superior average percent correct classification performance %over those corresponding to the $T_{P}$ and $TF_{FE}$ scenarios 
for SNR values of 18~{dB} and lower. The CGAN approach achieves the best average percent correct classification performance for all SNR values with a maximum of 99.85\% at an SNR of 30~{dB} and a minimum of 95\% at an SNR of 9~{dB}. Based on the results in Fig.~\ref{fig:DL_Trad_Comp}, DL-based equalization using the semi-supervised learning approaches presented in Sect.~\ref{sect:methodology} provides a better mechanism for compensating for multipath channel effects while simultaneously preserving SEI features.

%Scalability Analysis
\subsection{Results for Experiment \#2}
\label{sec:results_exp02}
\begin{figure}[!b]
	\centering
	\vspace{-5mm}
	\includegraphics[width=0.45\textwidth]{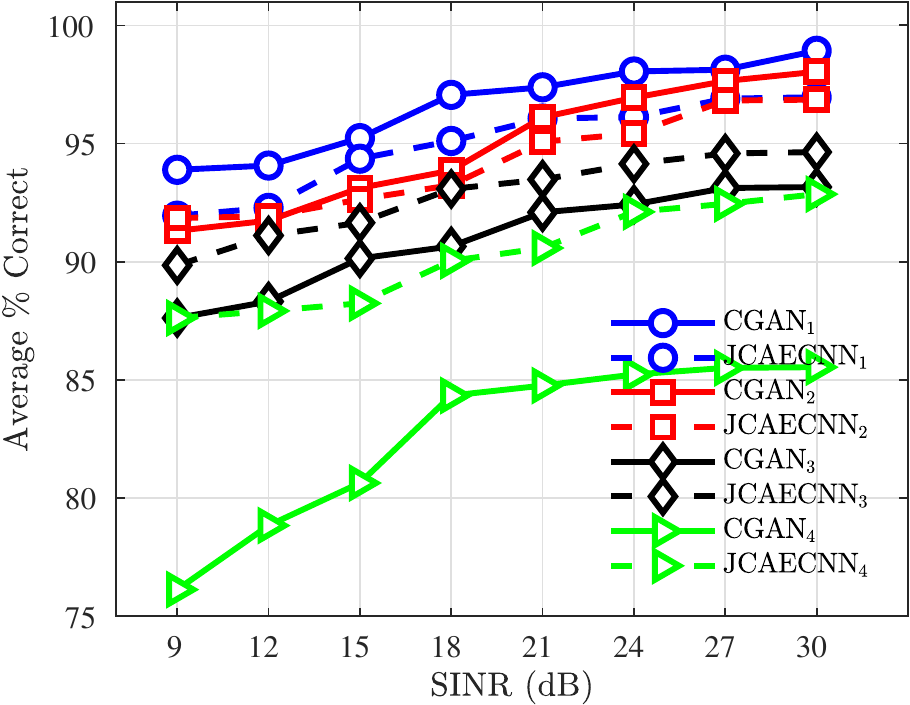}
	\caption{\underline{\textit{Experiment \#2 Results:}} Average percent correct classification performance across up to  $N_{D} = 32$ IEEE 802.11a TP-Link adapters using CGAN approach, JCAECNN, for SNR$\in$[9, 30]~{dB} in steps of 3~{dB}. Note that subscripts denote scenario number.}
	\label{fig:DL_scalability}
% 	\vspace{-5mm}
\end{figure}

The CGAN and JCAECNN equalization and classification approaches are assessed under an increased number of emitters to reflect larger IoT device deployments. The number of to be identified emitters is: % number of emitters to be identified in which there are: 
$N_{D} = [4, 8, 16, 32]$. %, $N_{D} = 8$, $N_{D} = 16$, or $N_{D} = 32$ emitters that need to be identified. 
This additional assessment is conducted using the preambles extracted from the signals transmitted by thirty-two TP-Link AC1300 USB Wi-Fi adapters. The assessment is conducted using the following four scenarios.
\begin{itemize}[leftmargin=*]
\item{\textit{Scenario \#1:} $N_{B}$ $=$ 10,000 preambles extracted from the signals transmitted by Emitter \#1 through Emitter \#4.}%collected from each emitter from \#1 to \#4.}
\item{\textit{Scenario \#2:} $N_{B}$ $=$ 10,000 preambles extracted from the signals transmitted by Emitter \#1 through Emitter \#8.}%preambles collected from each emitter from \#1 to \#8.}
\item{\textit{Scenario \#3:} $N_{B}$ $=$ 10,000 preambles extracted from the signals transmitted by Emitter \#1 through Emitter \#16.}%preambles collected from each emitter from \#1 to \#16.}
\item{\textit{Scenario \#4:} $N_{B}$ $=$ 10,000 preambles extracted from the signals transmitted by Emitter \#1 through Emitter \#32.}%preambles collected from each emitter from \#1 to \#32.}
\end{itemize}
For each scenario, a unique $L = 5$ Rayleigh fading channel is convolved with each collected preamble prior to adding scaled and like-filtered AWGN to achieve SNR values of 9~{dB} to 30~{dB} with 3~{dB} steps between consecutive SNR values and ten noise realizations per SNR. CGAN and JCAECNN approaches equalize and classify the preambles of each scenario using the same procedure described earlier in this section, but with $N_{R} = 8,000$ and $N_{T} = 2,000$ for each SNR, noise realization, and emitter. The average percent correct classification of the CGAN and JCAECNN approaches for each of the four scenarios is shown in Fig.~\ref{fig:DL_scalability} where CGAN$_{i}$ and JCAECNN$_{i}$ denote the CGAN and JCAECNN performance results for the $i^{\text{th}}$ scenario, respectively. \\%The subscript $i$ is used to denote the preambles of the $i^{\text{th}}$ group.
\indent Fig.~\ref{fig:DL_scalability} shows the average percent correct classification results for all four scenarios of Experiment \#2. When discriminating four emitters, % are used, the two 
both approaches result in similar performance as that shown in %show the same performance result as that in 
Fig.~\ref{fig:DL_Trad_Comp} with the CGAN approach achieving superior average percent correct classification performance for all SNRs. As the number of to be identified emitters increases, the performance of the JCAECNN approach begins to surpass that of the CGAN approach. When a total of sixteen emitters are used (a.k.a., Scenario \#3), the JCAECNN performance exceeds that of the CGAN approach by a minimum of 1\% for all SNRs. When the entire set of thirty-two emitters is used (a.k.a., Scenario \#4), the JCAECNN exceeds the performance of the CGAN approach for all SNRs. The smallest  improvement is 5\% at an SNR of {18}~{dB} and the largest improvement is 11\% at an SNR of {9}~{dB}.\\
\indent Fig.~\ref{fig:DL_scalability} shows that JCAECNN performance % the performance of the  approach 
suffers less degradation as the number of to be identified emitters %being identified 
increases from four to thirty-two; thus, this approach scales better for larger IoT deployments. The poorer performance of the CGAN approach is attributed to the increasing number of conditional distributions that the $G$ must learn %be learned by the $G$ 
as the number of emitters goes from four to thirty-two. This is exacerbated by the decreasing efficiency of the hidden label representation. 
In terms of decision complexity, the JCAECNN performs a total of $L+1$ CNN decisions corresponding to the reconstructed $\hat{C}_{k}$ and the CNN classifier output $\hat{y}_{c}$ for each received preamble $r[n]$. The number of JCAECNN classifications does not depend on the number of emitters (i.e., labels), which is not the case for the CGAN. In the CGAN approach, assigning preamble $r[n]$ to a label $y$ requires $N_{D}$ total equalization actions by the $G$ that is followed by an additional $N_{D}$ sequential classifications. This is to due the fact that the identity of the transmitting emitter is unknown; thus, the preamble must be compared to every known emitter's class. So, as the number of emitters linearly increases so does the number of conditional distributions learned by the $G$ as well as the number CGAN equalizations and classification decisions.% increases linearly.
 
%
%JCAECNN: Different Data Sets
\subsection{Results for Experiment \#3}
\label{sec:results_exp03}
\begin{figure}[!t]
	\centering
% 	\vspace{-5mm}
	\includegraphics[width=0.45\textwidth]{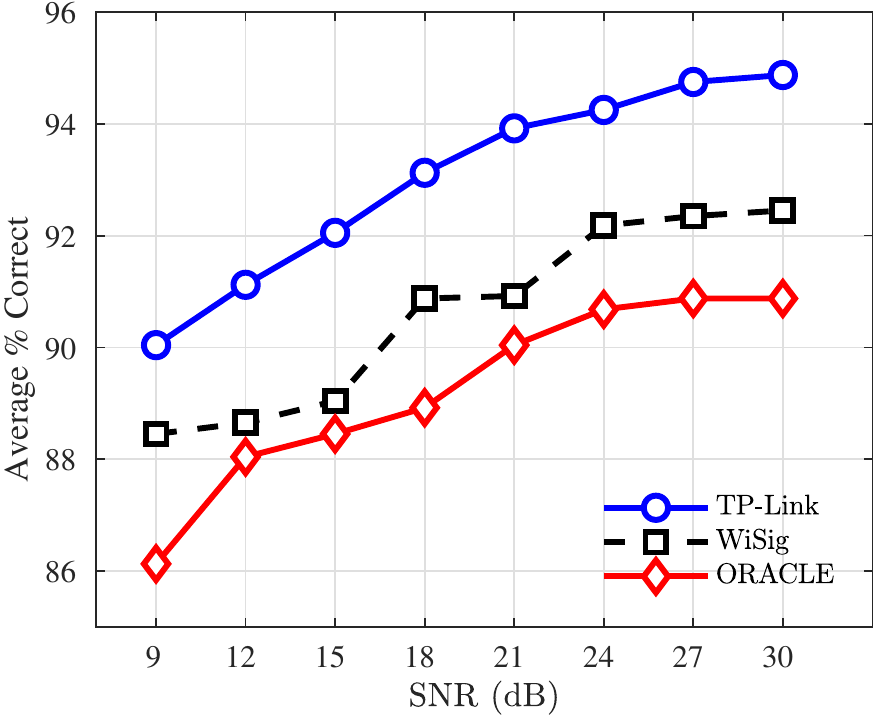}
	\caption{\underline{\textit{Experiment \#3 Results:}} Average percent correct classification performance for the TP-Link, WiSig, and ORACLE data sets--that each contain signals collected from $N_{D} = 16$ IEEE 80d2.11a Wi-Fi emitters--using the JCAECNN approach for
	SNR$\in$[9, 30]~{dB} in steps of 3~{dB} between consecutive values.}
	\label{fig:multi_data}
	\bigskip
	\vspace{-5mm}
\end{figure}

Based upon the results presented in Sect.~\ref{sec:results_exp02}, the results in this section are generated using only the JCAECNN approach. Fig.~\ref{fig:multi_data} shows the average percent correct classification performance of the JCAECNN approach for the following three different data sets that each contain signals collected from at least sixteen IEEE 802.11a Wi-Fi compliant emitters.
\begin{itemize}[leftmargin=*]
\item{\textit{Data Set \#1:} This data set consists of the IEEE 802.11a Wi-Fi preambles used to generate the Scenario \#3 results presented in Fig.~\ref{fig:DL_scalability} and explained in Sect.~\ref{sec:results_exp03}. This data set's results are designated using ``TP-Link''.}
\item{\textit{Data Set \#2:} This data set was collected by the authors of~\cite{Oracle_2019} to evaluate the Optimized Radio clAssification through Convolutional neuraL nEtworks (ORACLE) approach. The data set contains IEEE 802.11a Wi-Fi signals collected from 16 USRP X310 Software-Defined Radios (SDRs) using a stationary Ettus Research USRP B210 SDR as the receiver. It is important to note that the B210 receiver's sampling rate was set to 5~{MHz}, which is lower than the 20~{MHz} sampling rate of the collected signals used to generate the results presented up to this point. This data set's results are designated using ``ORACLE''.%The collected IEEE 802.11a signals are sampled at rate of 5 MHz.
}
\item{\textit{Data Set \#3:} This data set was collected by the authors of~\cite{WiSig} and designated the WiFi Signal (WiSig) data set. The WiSig data set contains IEEE 802.11a Wi-Fi signals captured from a total of 174 transmitting emitters--including USRP B210, X310, and N210 SDRs--over a four day period and using forty-one USRP receivers. %The data set contains 802.11 signals collected on four different days and using a total of 41 USRP receivers. 
In order to maintain consistency with the other two data sets and the rest of this paper's results only a single day's and receiver's worth of WiSig signals are used. %For the sake of consistency with the rest of the experiments in this work, only a portion of WiSig data set containing a single-day signals collected by a single receiver is considered in this work.
The chosen WiSig signals are detected using auto-correlation performed using the preamble's STS portion %of the preamble 
and then re-sampled to rate of 20~{MHz}. This data set's results are designated using ``WiSig''.
}
\end{itemize}
Similar to the TP-Link data of Scenario~{\#3}, each signal in the Oracle and WiSig data sets is filtered using a unique $L = 5$ Rayleigh fading channel prior to adding scaled and like-filtered AWGN to achieve SNR values of 9~{dB} up to 30~{dB} with 3~{dB} steps between consecutive SNR values and ten noise realizations per SNR. After that, the JCAECNN %joint CAE \& CNN-based 
approach performs equalization and classification as described earlier in this section. For the sake of consistency, a total of sixteen emitters are randomly selected from the WiSig data set to match the number of emitters represented in the TP-Link and ORACLE data sets. For TP-Link and Oracle data sets, $N_{R} = 8,000$ and  $N_{T} = 2,000$. For the single day and receiver portion of the WiSig data, the total number of signals collected per emitter is 1,000; thus, $N_{R} = 800$ and  $N_{T} = 200$ to provide a ratio consistent with the other two data sets. %Using the same ratio, the $N_{B_{R}}$ and $N_{B_{T}}$ parameters of the WiSig data set are set to 800 and 200, respectively. 
The average percent correct classification results presented in Fig.~\ref{fig:multi_data} show that %the average percent correct classification of the 
JCAECNN equalization and classification of the WiSig preambles achieves superior performance to that of the Oracle data for all SNRs despite the limited number of training samples. That can be attributed to the lower sampling rate used to collect the ORACLE data, which represents only 25\% of the sampling rate of the TP-Link and ORACLE data. Additionally, the WiSig data set is comprised of multiple USRP models while the ORACLE signals are transmitted by USRPs that are all the same model. The latter is serial number discrimination, which is the most challenging SEI case. Additionally, ORACLE's USRP X310 is a \$8,500 high-performance SDR made with low variability components, which means less feature variability across SDRs making SEI more difficult~\cite{Oracle_2019}.
%
%JCAECNN Loss Optimization
\subsection{Results for Experiment \#4}
\label{sec:results_exp04}
\begin{figure*}[!t]
    \centering
    \subfigure[$N_{D} = 4$ emitters.]{\label{loss_weights_a}\includegraphics[width=0.45\textwidth]{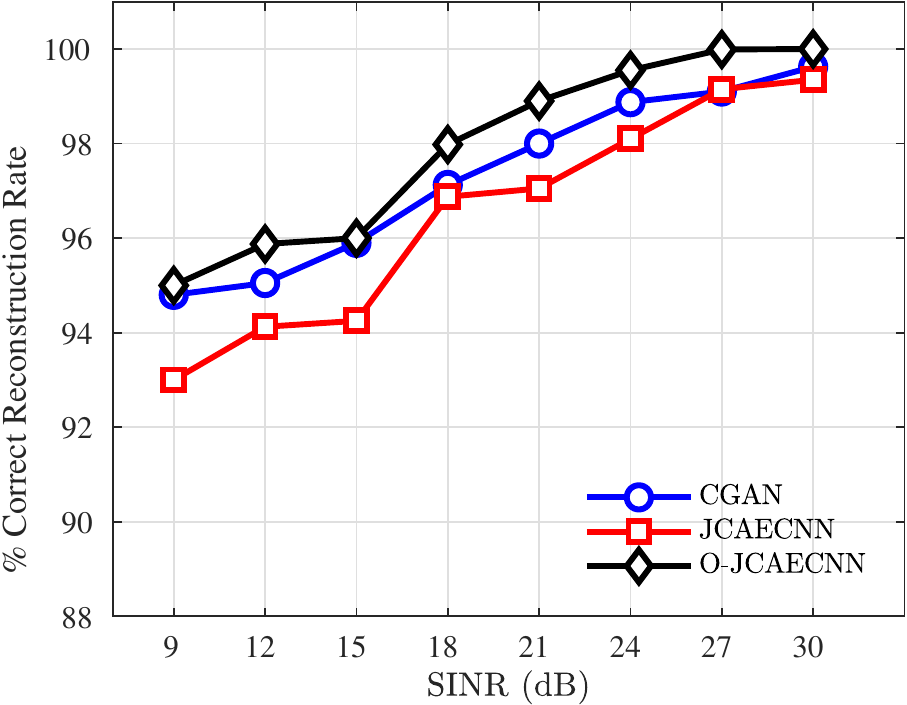}}
    ~
    \subfigure[$N_{D} = 16$ emitters.]{\label{fig:loss_weights_b}\includegraphics[width=0.45\textwidth]{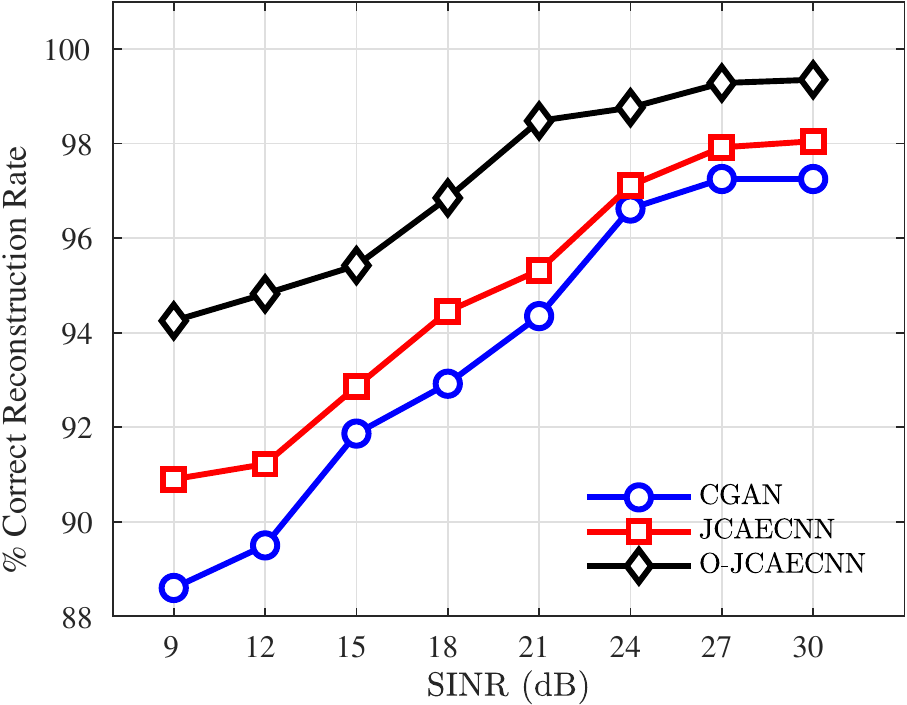}}
    \caption{\underline{\textit{Experiment \#4 Results:}} Average percent correct classification performance across $N_{D}$ IEEE 802.11a Wi-Fi TP-Link Adapters using the CGAN, JCAECNN, and O-JCAECNN approaches for SNR$\in$[9, 30]~{dB} in steps of 3~{dB}.}
    \label{fig:loss_weights}
    \vspace{-5mm}
\end{figure*}
Up to this point, all results generated by the JCAECNN approach set $\lambda_{k}$ and $\lambda_{c}$ to one. In an effort to optimize $L=5$ Rayleigh fading channel equalization and classification performance %the equalization and classification performance for an $L=5$ Rayleigh fading channel, 
the loss function weights $\lambda_{k}$ for $k = [1, 2,3,4,5]$ %$\lambda_{2}$, $\lambda{3}$, $\lambda{4}$, $\lambda{5}$, 
and $\lambda_{c}$ are set to: thirty-two, sixteen, eight, four, two, and thirty-two, respectively. Selection of these values is based on the fact that the variances of the corresponding Rayleigh channel paths given in equation \eqref{eqn:variance} decay exponentially with $k$. Fig.~\ref{fig:loss_weights} shows the average percent correct classification for the JCAECNN approach with decaying loss weights and is designated as %are used. These results are denoted by 
O-JCAECNN in which the `O' denotes `optimized'. Comparative assessment is conducted using %enabled by including performance 
results generated by: (i) the JCAECNN implementation in which the loss weights are the same (i.e., they are all set to a value of one) % and designated as \hl{JCAECNN$_{S}$} 
and (ii) the CGAN results presented in Fig.~\ref{fig:DL_Trad_Comp}. 

\begin{figure}[!b]
	\centering
	\vspace{-5mm}
	\includegraphics[width=0.45\textwidth]{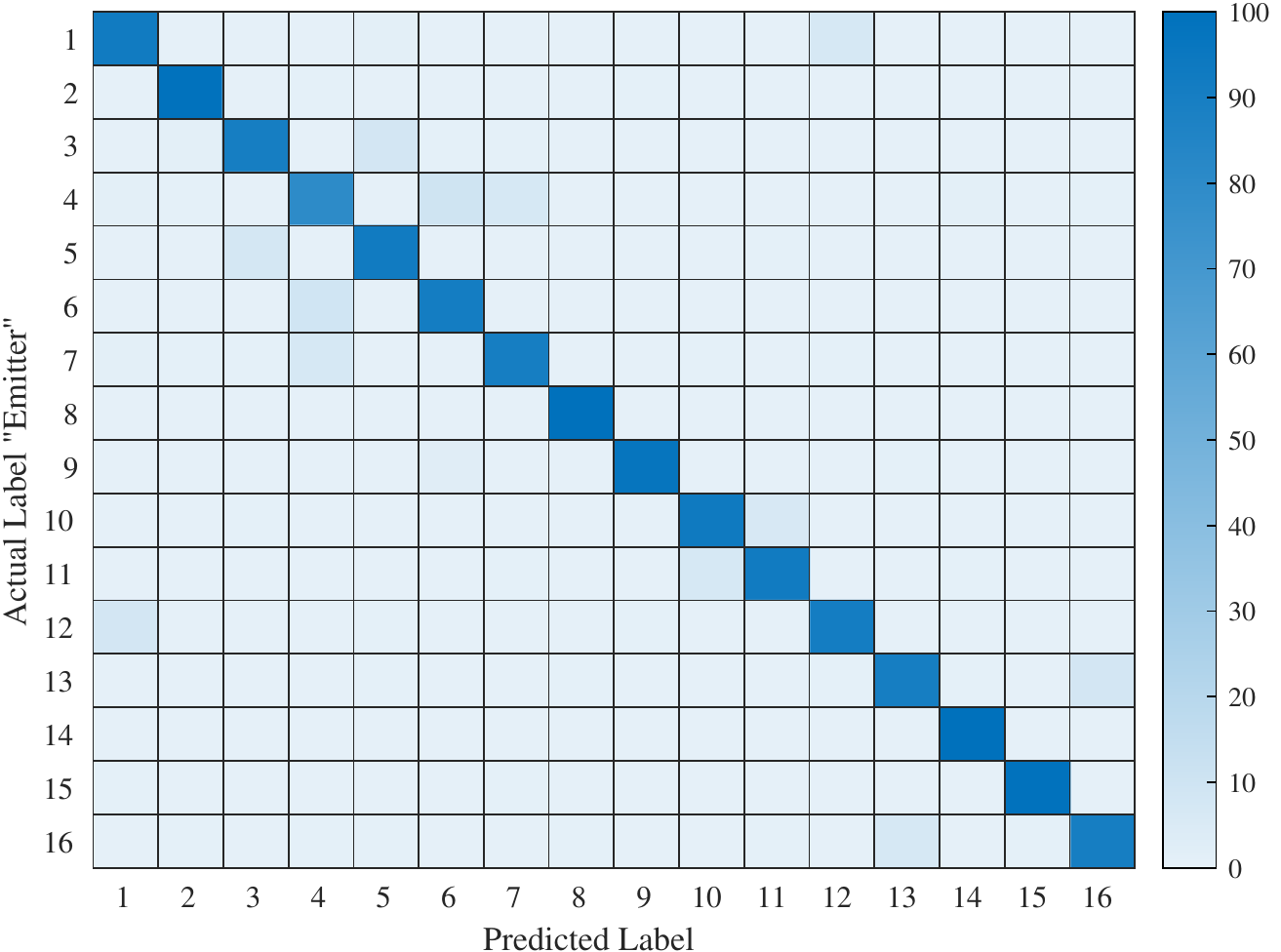}
	\caption{\underline{\textit{Experiment \#4 Results:}} Confusion matrix showing individual percent correct classification performance for each of the  $N_{D}=16$ IEEE 802.11a Wi-Fi TP-Link Adapters provided by Scenario \#3 in Sect.~\ref{sec:results_exp02} using the JCAECNN approach for SNR $=$ 9~{dB}. Average percent correct classification performance is 94.25\%.}
	\label{fig:Confuion_matrix}
% 	\vspace{-5mm}
\end{figure}

Superior average percent correct classification performance is achieved using the O-JCAECNN approach %with exponentially decaying loss function weights 
for all SNR values of 9~{dB} to 30~{dB} using a 3~{dB} step between consecutive values. It is the only SEI approach to achieve an average percent correct classification performance of 100\% for $N_{D} = 4$ case, which occurs at an SNR of 27~{dB} and 30~{dB}. Fig.~\ref{fig:Confuion_matrix} is the confusion matrix corresponding to the 9~{dB} O-JCAECNN results presented in Fig.~\ref{fig:loss_weights_b} and is included to show percent correct classification performance for each of the $N_{D} = 16$ emitters. It can be seen %from Fig.~\ref{fig:Confuion_matrix} 
that eleven of the sixteen emitters %except emitter \#3, \#4, \#5, \#6, and \#12 
are represented by dark blue diagonal entries, which indicates a percent correct classification performance of at least 95\%. The exceptions being Emitter~{\#3}, Emitter~{\#4}, Emitter~{\#5}, Emitter~{\#6}, and Emitter~{\#12} of which only Emitter~{\#4} is not classified correctly at least 90\% of the time.% All emitters except emitter \#4 achieve at least  90\% correct classification performance.

The results show %of this experiment show Based on the observation and results of this experiment, we can say that 
that the performance of the JCAECNN model is dominated by: (i) the loss weights in equation \eqref{eqn:custom_loss} that correspond to the first few Rayleigh fading paths, and (ii) the loss weight $\lambda_{c}$ corresponding to the CNN$_{I}$ classifier head. This suggests that channel-invariant RF-DNA fingerprinting is possible by optimizing the first few and most dominant weights of the JCAECNN's loss function. %This sets the basis for the design of a channel-invariant RF-DNA fingerprinting system as the equalization error due to the assumption of less number of Rayleigh channel paths can be reduced by optimizing the loss weights of the first few and more significant terms of the loss function.
%
\begin{comment}
\begin{figure}[!t]
	\centering
	%\vspace{8mm}
    \includegraphics[width=0.45\textwidth]{figures/eps/loss_weights.eps}
    \caption{\underline{\textit{Experiment \#4 Results:}} Average percent correct classification performance across   $N_{D}=4$ IEEE 802.11a Wi-Fi emitters using CGAN approach, JCAECNN 'CAE$_{s}$', and JCAECNN 'CAE$_{D}$'
    for SNR$\in$[9, 30]~{dB} in steps of 3~{dB}}
    \label{fig:loss_weights}
    \vspace{-5mm}
\end{figure}

\begin{figure}[!t]
	\centering
	%\vspace{8mm}
    \includegraphics[width=0.45\textwidth]{figures/eps/loss_weights_16.eps}
    \caption{\underline{\textit{Experiment \#4 Results:}} Average percent correct classification performance across   $N_{D}=16$ IEEE 802.11a Wi-Fi TP-Link Adapters using CGAN approach, JCAECNN, and Optimized JCAECNN for SNR$\in$[9, 30]~{dB} in steps of 3~{dB}}
    \label{fig:loss_weights}
    \vspace{-5mm}
\end{figure}
\end{comment}
%
\section{Conclusion and Future Work}%
\label{sect:SumConc}
This work presents and analyzes two semi-supervised DL approaches--designated herein as CGAN and JCAECNN--to enhance IoT security using SEI under Rayleigh fading and degrading SNR. %by significantly improving the SEI performance under Rayleigh fading channels and degrading SNR. 
These approaches are capable of extracting discriminating RF-DNA fingerprint features from signals transmitted by up to thirty-two IEEE 802.11a Wi-Fi emitters while traversing a %and undergone 
$L = 5$ Rayleigh fading channel. The two architectures %based on CGAN and joint CAE \& CNN (JCAECNN) 
can reconstruct the transmitted preamble from its received, multipath corrupted version while preserving the SEI discriminating features using the SNR robust, IQ+NL preamble representation. %s to represent each preamble due to it's robustness against noise. 
The two semi-supervised learning approaches are compared with the traditional RF-DNA fingerprinting approaches in~\cite{Fadul_Access_2021,Fadul_GIoTS_2022}. %with the traditional, equalization-based RF-DNA fingerprinting given by PTRFF and TFRFF. 
The JCAECNN approach shows better scalability as the number of emitters increases from $N_{D} = 4$ to $N_{D} = 32$, which is not the case for the CGAN approach due to the number of preamble reconstructions and classification decisions increasing linearly with the number of emitters. %.Due to the fact that the number of preamble reconstructions and classification decisions performed by the CGAN approach is linearly dependent on the number of emitters,  
The O-JCAECNN results in superior average percent correct classification performance that exceeds 94\% for sixteen IEEE 802.11a Wi-Fi emitters at SNR values of 9~{dB} and higher.\\ %improves the performance of Pre-trained CNN RF fingerprinting (PTRFF) and TF Feature-engineered RF-DNA fingerprinting (TFRFF) by XX\% and YY\% respectively}.
\indent Future research is focused on increasing the viability of SEI-based IoT security by modifying the CGAN and JCAECNN architectures to: (i) allow detection of emitters that are not represented within the training signals set, and (ii) leverage simultaneous training by integrating the collaborative learning scheme presented in~\cite{song2018collaborative}.
~\cite{sutton2018reinforcement}
%Modify the existing CGAN- and joint CAE \& CNN-based designs to provide intrusion detection by identifying emitters not seen during training. That will enhance the IoTUse the collaborative learning scheme presented in~\cite{song2018collaborative} to reduce the computational complexity of joint CAE \& CNN by enabling all decoder heads to train simultaneously instead of sequentially.
%
Goldsmith, A. (2005). Wireless Communications. Cambridge: Cambridge University Press. doi:10.1017/CBO9780511841224

\balance

\bibliographystyle{IEEEtran}

\bibliography{InfoSec2022_bib_v01} \clearpage

\end{document}